\documentclass[useAMS,usenatbib]{mn2e}

\usepackage[active]{srcltx}
\usepackage{graphicx}
\usepackage{txfonts}
\usepackage{natbib}
\usepackage{multirow}
\usepackage{longtable}
\usepackage[applemac]{inputenc}
\usepackage{rotating}
\usepackage{bigdelim}
\usepackage{subfig}

\usepackage{color}
\usepackage{soul}
\definecolor{jaune}{rgb}{1.0, 1.0, 0.0}

\newcommand{\kms}{km\,s$^{-1}\,$}

\newcommand{\bz}{\ensuremath{\langle B_z\rangle}}
\newcommand{\nz}{\ensuremath{\langle N_z\rangle}}

\def\gtrsim{\mathrel{\hbox{\rlap{\hbox{\lower4pt\hbox{$\sim$}}}\hbox{$>$}}}}
\def\ltsim{\mathrel{\hbox{\rlap{\hbox{\lower4pt\hbox{$\sim$}}}\hbox{$<$}}}}

\title[Weak magnetic fields in $\iota$~Herculis]{A search for weak or complex magnetic fields in the the B3V star $\iota$~Herculis \thanks{Based on observations obtained at the Canada-France-Hawaii Telescope (CFHT) which is operated by the National Research Council of Canada, the Institut National des Sciences de l'Univers (INSU) of the Centre National de la Recherche Scientifique of France, and the University of Hawaii.}}
\author[G.A. Wade et al.]{G.A. Wade\thanks{E-mail: wade-g@rmc.ca}$^1$, C.P. Folsom$^2$, P. Petit$^2$, V. Petit$^3$, F. Ligni\`eres$^2$, M. Auri\`ere$^2$,  T. B\"ohm$^2$
\\
$^{1}$Dept. of Physics, Royal Military College of Canada, PO Box 17000 Station Forces, Kingston, ON, Canada K7K 0C6 \\
$^{2}$IRAP - CNRS/Universit\'e de Toulouse, 14 Av. E. Belin, Toulouse, F-31400, France\\
$^3$Department of Physics and Astronomy, Bartol Research Institute, University of Delaware, Newark, DE 19716, USA\\
}

\begin{document}

\date{Accepted . Received , in original form }

\pagerange{\pageref{firstpage}--\pageref{lastpage}} \pubyear{2002}

\maketitle

\label{firstpage}

\begin{abstract}

We obtained 128 high signal-to-noise ratio Stokes $V$ spectra of the B3V star $\iota$~Her on 5 consecutive nights in 2012 with the ESPaDOnS spectropolarimeter at the Canada-France-Hawaii Telescope, with the aim of searching for the presence of weak and/or complex magnetic fields. Least-Squares Deconvolution mean profiles were computed from individual spectra, averaged over individual nights and over the entire run. No Zeeman signatures are detected in any of the profiles. The longitudinal magnetic field in the grand average profile was measured to be $-0.24 \pm	0.32	$~G, as compared to $ -0.22 \pm 0.32$~G in the null profile. Our observations therefore provide no evidence for the presence of Zeeman signatures analogous to those observed in the A0V star Vega by \citet{2009A&A...500L..41L}. We interpret these observations in three ways. First, we compare the LSD profiles with synthetic Stokes $V$ profiles corresponding to organised (dipolar) magnetic fields, for which we find an upper limit of about 8~G on the polar strength of any surface dipole present. Secondly, we compare the grand average profile with calculations corresponding to the random magnetic spot topologies of \citet{2013A&A...554A..93K}, inferring that spots, if present, of 2 degrees radius with strengths of 2-4 G and a filling factor of 50\% should have been detected in our data. Finally, we compare the observations with synthetic $V$ profiles corresponding to the surface magnetic maps of Vega \citep{2010A&A...523A..41P} computed for the spectral characteristics of $\iota$~Her. We conclude that while it is unlikely we would have detected a magnetic field identical to Vega's, we would have likely detected one with a peak strength of about 30 G, i.e. approximately four times as strong as that of Vega.

\end{abstract}



\section{Introduction}



%



Stellar magnetic fields contribute in an important way to the structure, dynamics and energetics of stellar atmospheres, envelopes and winds. In cool stars with convective envelopes (spectral types F, G, K and M), magnetic fields are produced through the dynamo processes acting in these stars. These often complex and variable magnetic fields are essentially ubiquitous, and exhibit characteristics that are strongly correlated with stellar convective and rotational properties \citep[e.g.][]{2013arXiv1309.6970W}. In contrast, magnetic fields of hotter stars with radiative envelopes (spectral types A, B and O) are characterised by simpler global topologies that are observed to be remarkably stable on long (years-decades) timescales. These magnetic fields are in fact quite rare: only a small fraction \citep[less than 10\%; e.g.][]{2008CoSka..38..443P, 2013arXiv1310.3965W} are known to host such fields, with the remainder being observably non-magnetic \citep[e.g.][]{2010A&A...523A..40A}. The existence of this 'magnetic dichotomy' \citep{2007A&A...475.1053A} is arguably the most fundamental outstanding problem in our understanding of the magnetism of early-type stars.

Recently, \citet{2009A&A...500L..41L} reported the detection of an extraordinarily weak Stokes $V$ signature in spectral lines of the bright A1V star Vega. They proposed that these signatures were the consequence of Zeeman effect, implying the presence of a very weak and complex magnetic field located in the visible photosphere of the star. \citet{2010A&A...523A..41P} confirmed this hypothesis, reproducing the signature at 4 different epochs using 2 different instruments, and interpreting the spectropolarimetric time series in terms of a surface magnetic field using Zeeman Doppler Imaging (ZDI). The results of this procedure support the view that Vega is a rapidly rotating ($P_{\rm rot}=0.73$~d) star seen nearly pole-on ($i=7\degr$). The reconstruction of the magnetic topology at two epochs revealed a magnetic region of radial field orientation, closely concentrated around the rotation pole. This polar feature is accompanied by a small number of magnetic patches at visible lower latitudes. No significant variability of the field structure was observed over a one-year time span.

As proposed by \citet{2009A&A...500L..41L} and \citet{2010A&A...523A..41P}, Vega may well be the first confirmed member of a much larger, as yet unexplored, class of weakly-magnetic early-type stars. As such, it may provide important clues for understanding the magnetic dichotomy. In this paper we test this proposal by searching for Vega-like Stokes $V$ signatures in a hotter early-type star, $\iota$~Herculis.

$\iota$ Her (HR 6588, HD 160762) is a bright ($V=3.80$) B3 IV star. Due to its brightness and sharp ($v\sin i=6$~km/s) lines, it has been frequently employed as a spectroscopic and abundance standard \citep[e.g.][]{2010A&A...515A..74L, 2012A&A...539A.143N}. Its atmospheric and physical parameters are both accurately and precisely known, and were most recently reported by \citet{2013A&A...550A..26N} and \citet{2012A&A...539A.143N}. These characteristics also make $\iota$~Her well-suited to high precision magnetometry. The properties most relevant to this study are summarised in Table 1.

$\iota$ Her is a Slowly Pulsating B-type (SPB) pulsator with a principal pulsation frequency of 0.28671~c\,d$^{-1}$ \citep[about 3.5~d;][]{2000A&A...362..189C}. The pulsation amplitude is significantly variable on long ($>1$~yr) timescales. In fact \citet{2009MNRAS.398.1339H} was unable to detect any $uvy$ photometric variability at the $\sim$mmag level, whereas the photometric amplitude in the Hipparcos $B_{\rm T}$ filter reported by \citet{2000A&A...362..189C} was in excess of 15 mmag. \citet{2000A&A...362..189C} also report variable radial velocity (RV) attributable to pulsation. The RV amplitude is also secularly variable.

$\iota$~Her is also the primary of a single-lined spectroscopic binary (SB1) system,  with a $\sim 113$~d orbital period \citep{2000A&A...362..189C}. Chapellier et al. note that assuming a mass of $\sim 7~M_\odot$ for the primary, the mass function implies an upper limit of $0.4~M_\odot$ for the companion. They suggested that the companion is a white dwarf, although it could also be an M dwarf.

\citet{2013A&A...551A..30B} find a rather significant difference between Si abundances measured using lines of Si~{\sc i} and {\sc ii} in the spectrum of $\iota$~Her - in fact the most significant difference in their sample. But they ascribe this to NLTE, noting that \citet{2012A&A...539A.143N} find a solar abundance of this element. \citet{2009A&A...503..973L} observed no direct evidence of atmospheric velocity fields in the line profiles. \citet{1958ApJS....3..141B} included $\iota$~Her in his list of stars 'showing little or no evidence of Zeeman effect'.

In this paper we employ an extensive new collection of contiguous high resolution, high signal-to-noise ratio (SNR) spectra to constrain the presence of magnetic field in the photosphere of $\iota$~Her. The data acquisition, reduction and characteristics are described in Sect. 2, along with the Least-Squares Deconvolution. Evaluation of the magnetic field detection and longitudinal field measurement is discussed in Sect. 3. In Sect. 4 we describe the modeling performed to interpret our null results, both in terms of organised dipolar field structures, random spot topologies, and a field analogous to that detected on Vega by \citet{2009A&A...500L..41L}.


\begin{figure}
\begin{centering}
\includegraphics[width=8cm]{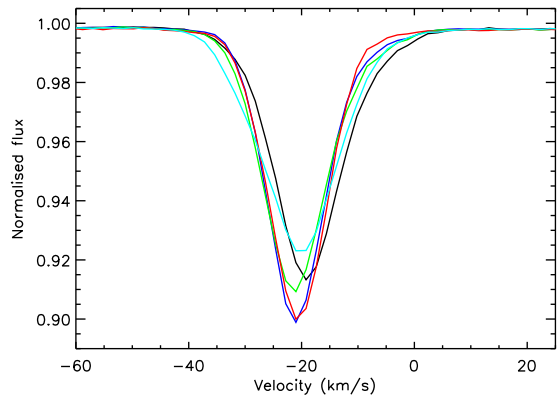}
\caption{\label{nightly}Mean nightly LSD profiles of $\iota$~Her. Night 1 - black. Night 2 - blue. Night 3 - green. Night 4 - red. Night 5 - cyan.}
\end{centering}
\end{figure}

%
%

\begin{table}
\caption{\label{params}$\iota$~Her physical and wind parameters. From \citet{2013A&A...550A..26N} and \citet{2012A&A...539A.143N} except $R/R_\odot$, which is computed from $M/M_\odot$ and $\log g$.}
\begin{tabular}{llllllllllllllrrc}
 Parameter  & Value \\
\hline 
$T_{\rm eff}$ (K) & $17500\pm 200$ \\ 
$\log g$ (cgs) & $3.80\pm 0.05$ \\
$M/M_\odot$ & $6.7\pm 0.2$ \\ 
$R/R_\odot$ &  $5.4\pm 0.2$\\
$v\sin i$ (km/s) & $6\pm 1$ \\
  \hline
\end{tabular}
\end{table}

\section{Observations}

One hundred twenty-eight Stokes $V$ sequences of $\iota$ Her were acquired with the ESPaDOnS spectropolarimeter \citep[see e.g.][for details and performance of ESPaDOnS]{2012MNRAS.426.1003S} mounted on the 3.6m Canada-France-Hawaii Telescope (CFHT) during 25-29 June 2012. Each polarimetric sequence consisted of 4 individual subexposures of 60s duration, between which the $\lambda/2$ retarders (Fresnel rhombs) are rotated. The data were acquired in fast readout mode. From each set of four subexposures we derived Stokes $I$ and Stokes $V$ spectra in the wavelength range 3670 - 10000~\AA\ following the double-ratio procedure described by \citet{1997MNRAS.291..658D} and \citet{2009PASP..121..993B}, which cancels spurious polarization signatures to first order. Diagnostic null polarization spectra $N$ were also calculated by combining the four subexposures in such a way that polarization cancels out, providing us with a second verification that no spurious signals are present in the data. All frames were processed using the automated reduction package Libre ESpRIT \citep{1997MNRAS.291..658D} fed by the Upena pipeline at CFHT. Continuum normalization was performed order-by-order, using an IDL tool developed by VP.

The median SNR of the reduced spectra is 1021 per 1.8 km/s pixel. The log of observations is reported in Table 2. The reduced data can be accessed at the CFHT archive hosted by the Canadian Astronomy Data Centre (CADC) accessible at http://www1.cadc-ccda.hia-iha.nrc-cnrc.gc.ca.

\subsection{Variability}

$\iota$~Her is an SPB star. Pulsations are reflected in its spectral lines through variable distortions and small RV shifts. This variability is detected in our spectra as evolving RVs (on the order of a few tenths of km/s) over the course of each observing night, and small differences in width, depth and RV of the nightly-averaged Least-Squares Deconvolved (LSD) line profiles (discussed further in the next section). To illustrate the magnitude of these effects we show the nightly-averaged LSD Stokes $I$ profiles in Fig.~\ref{nightly}.

The pulsation period of $\iota$~Her is rather long (3.5 d) compared to the exposure time (60s) and polarimetric sequence duration (370 s including readout). As a consequence, we do not observe any indication of spurious signatures introduced into the individual Stokes $V$ or null $N$ LSD profiles due to variability during the exposure sequences \citep[e.g.][]{2012A&A...546A..44N}. On the other hand, night-to-night variability of the line profile is significant  \citep[although very small compared to that of $\beta$~Cep pulsators like $\gamma$~Peg; ][]{2014A&A...562A..59N}. 

The projected rotational velocity $v\sin i$ of $\iota$~Her is well established, and equal to $6\pm 1$~km/s \citep{2013A&A...550A..26N}. Using the radius of $5.4\pm 0.2~R_\odot$ computed from the measured mass and surface gravity and cited in Table~\ref{params}, we can estimate the stellar rotational period assuming rigid rotation. The rotation period is given by $P_{\rm rot}=50.6\, R\sin i/v\sin i=(45.5_{-8}^{+11})\sin i$~d. If $\sin i=1.0$ then $P_{\rm rot}=37.5-56.5$~d. If $\sin i=0.7$ (i.e. $i\simeq 45\degr$), then $P_{\rm rot}=26.3-39.6$~d. If $\sin i=0.2$ (i.e. $i\simeq 12\degr$), then $P_{\rm rot}=7.5-11.3$~d. 

Previous asteroseismic analyses of $\iota$ Her \citep{1995A&A...300..200M, 2000A&A...362..189C} provide no guidance concerning neither the star's inclination nor its rotational period. 

\begin{figure*}
\begin{centering}
\subfloat[][Grand average]{\includegraphics[width=5.7cm]{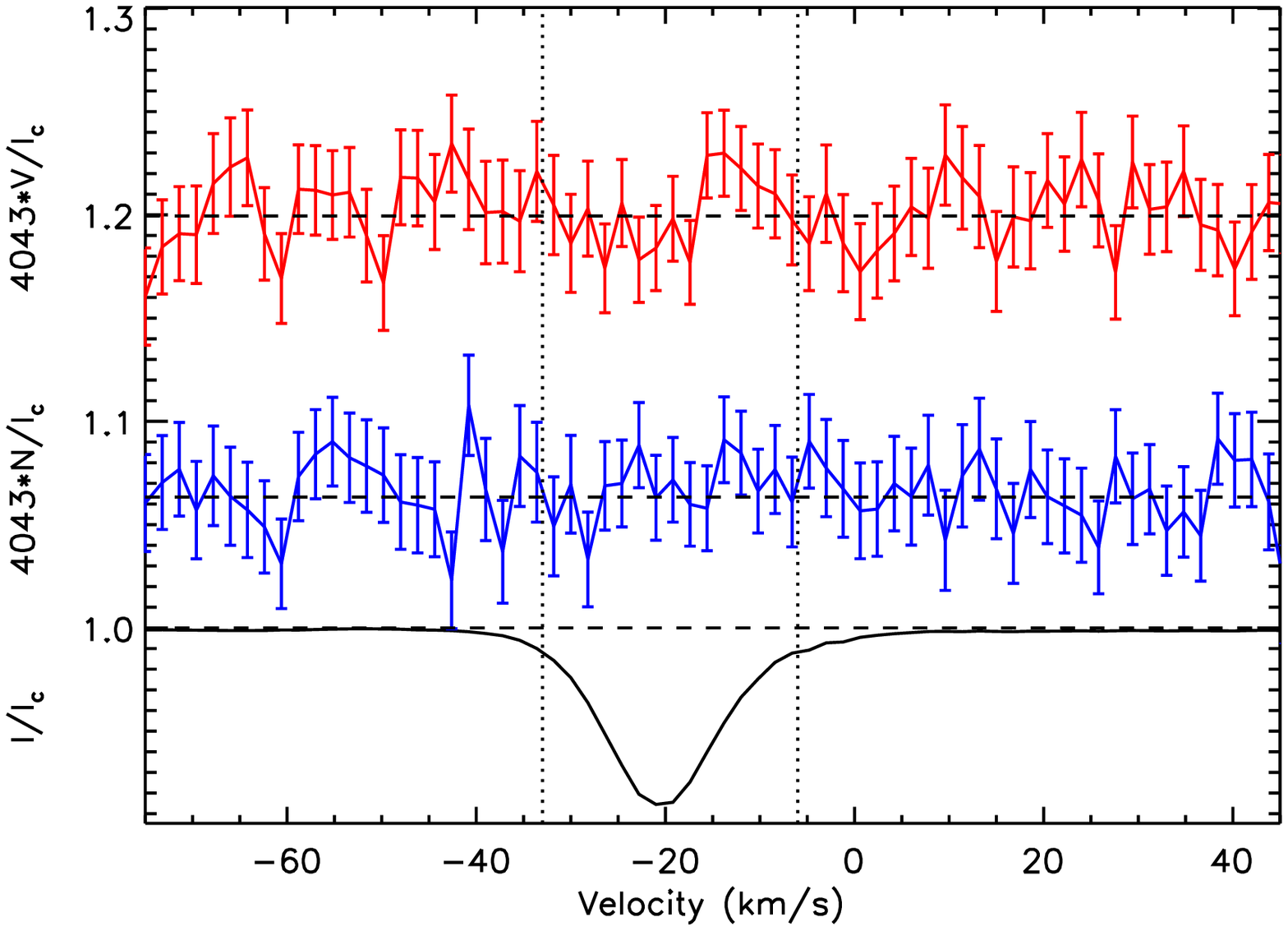}}\subfloat[][Night 1]{\includegraphics[width=5.7cm]{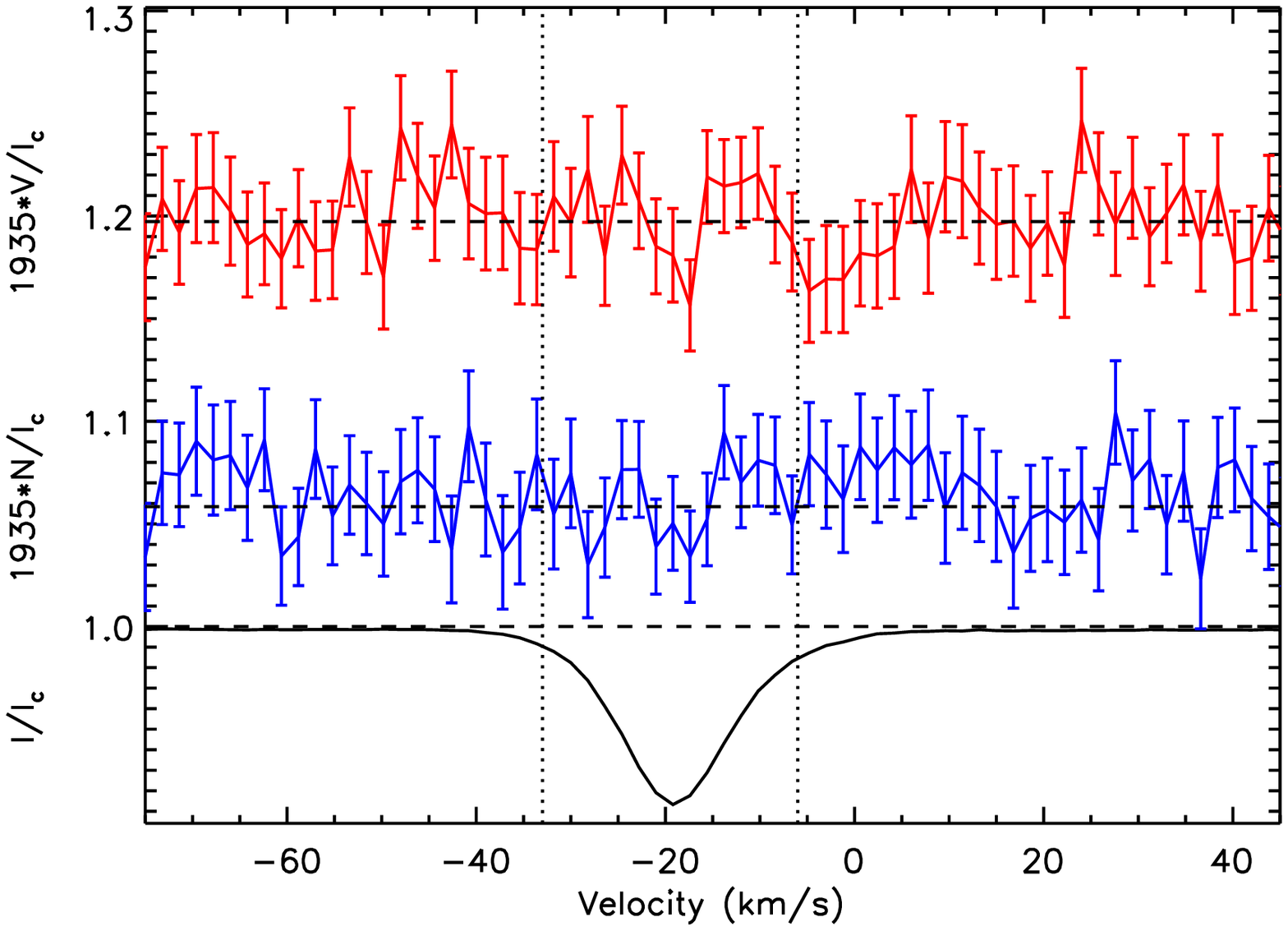}}\subfloat[][Night 2]{\includegraphics[width=5.7cm]{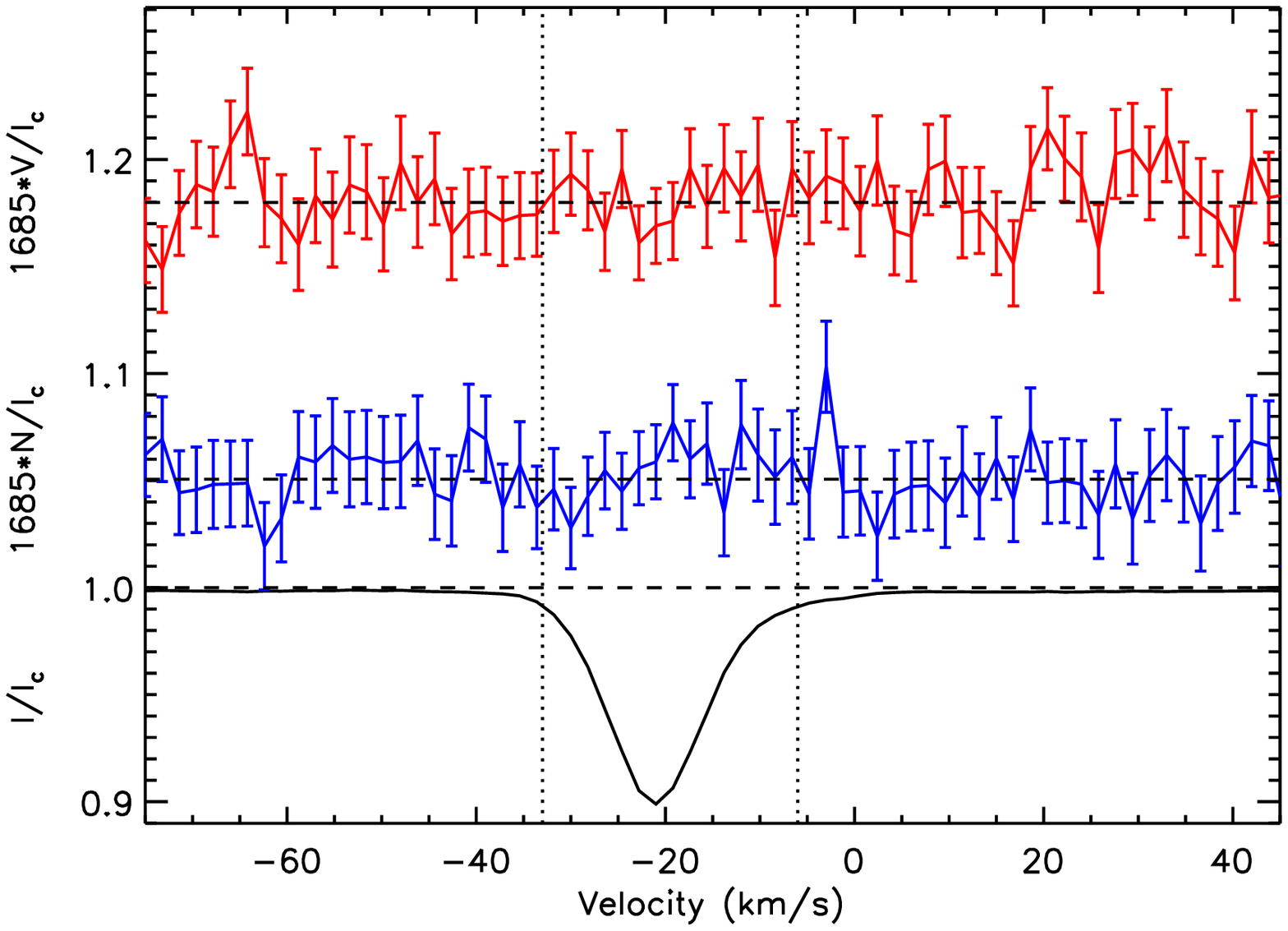}}\\
\subfloat[][Night 3]{\includegraphics[width=5.7cm]{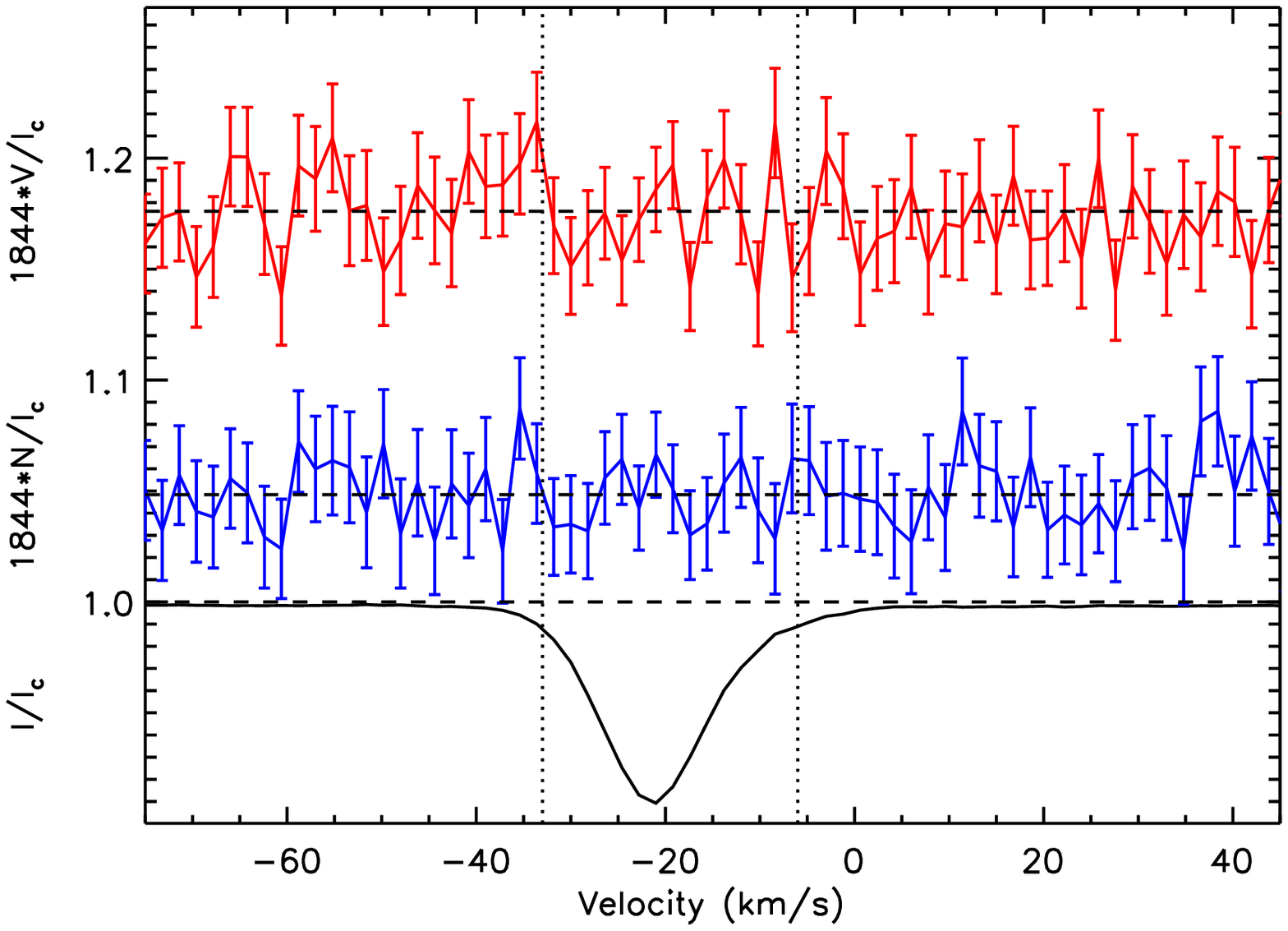}}\subfloat[][Night 4]{\includegraphics[width=5.7cm]{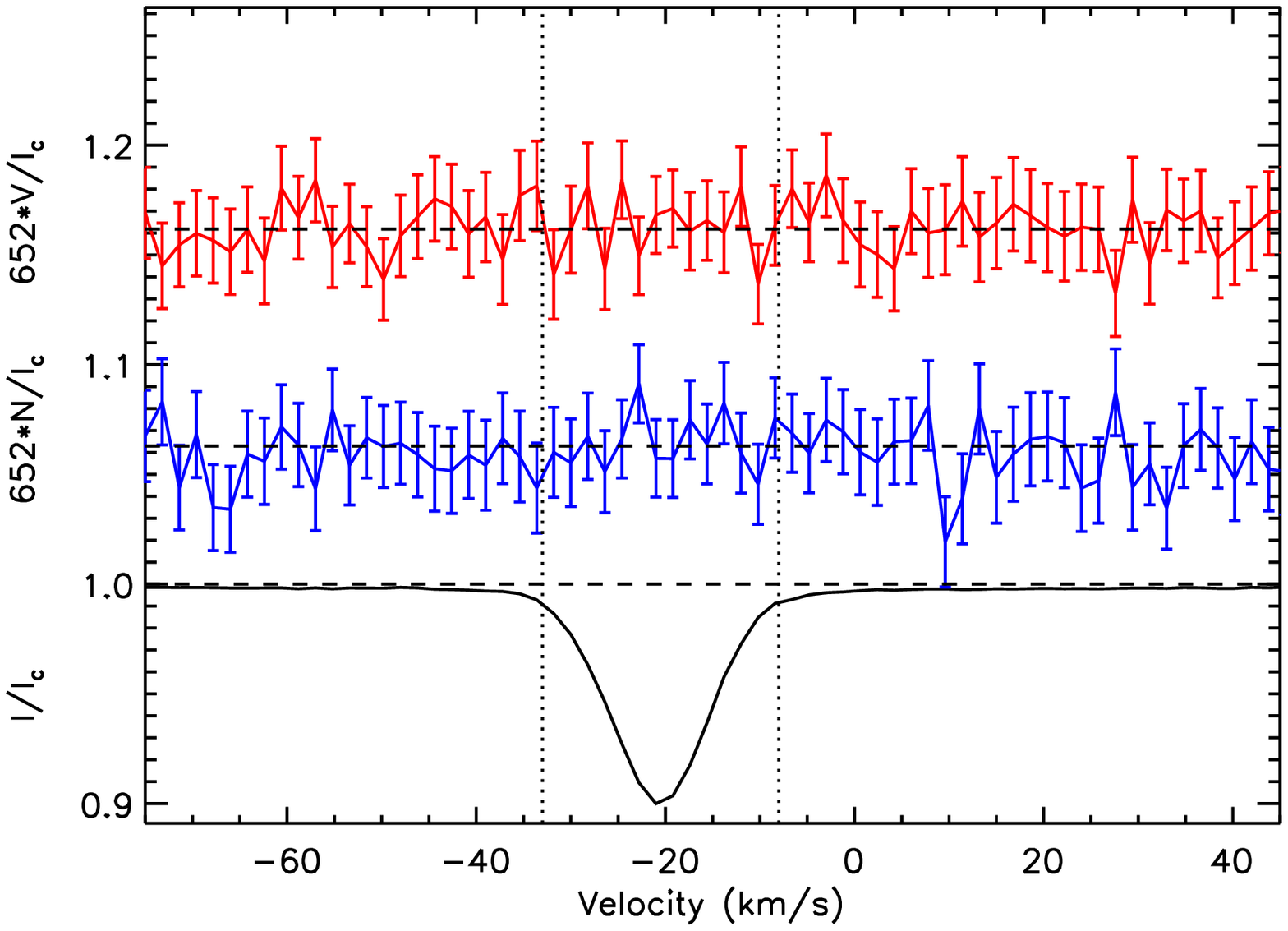}}\subfloat[][Night 5]{\includegraphics[width=5.7cm]{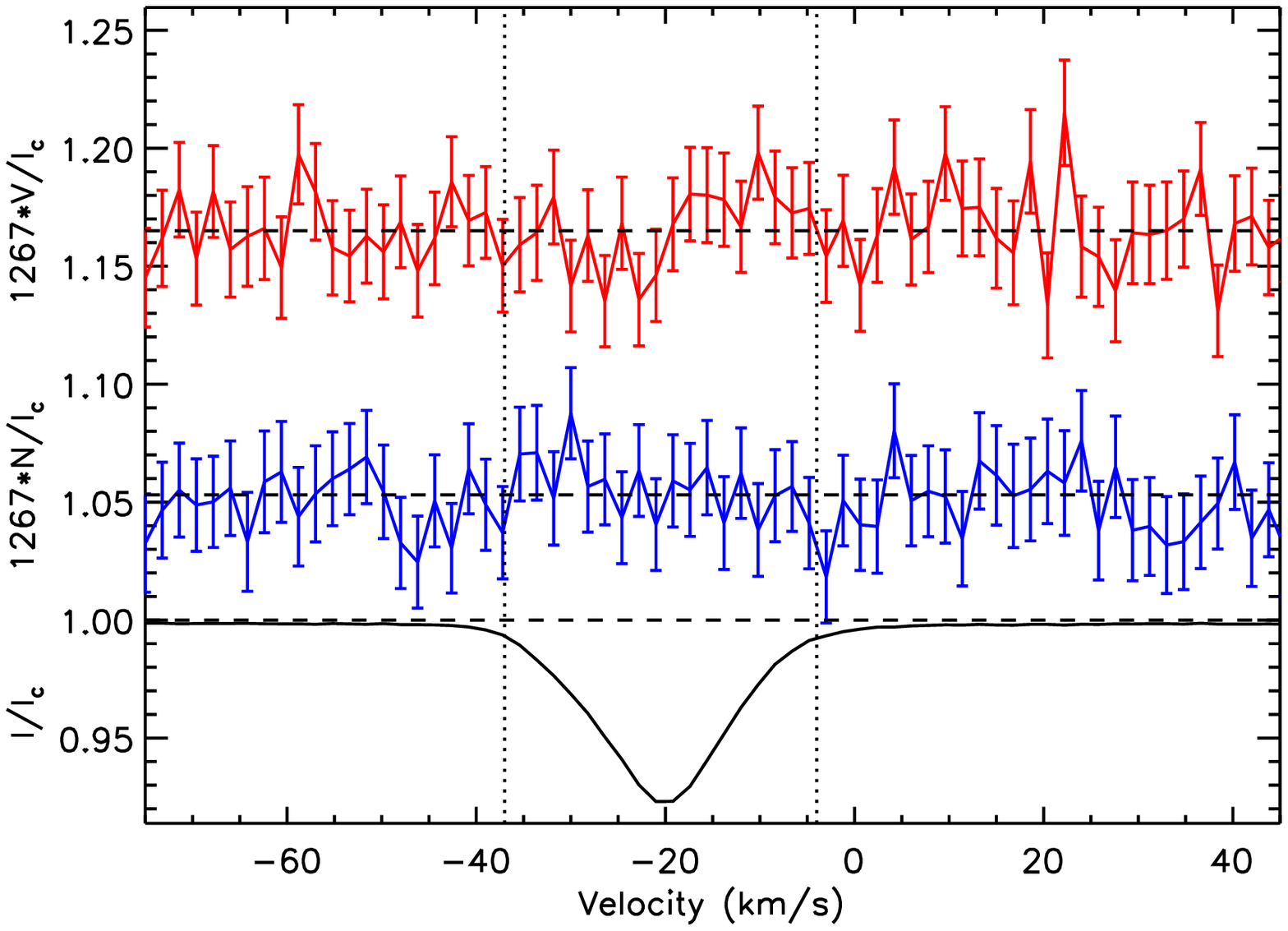}}
\caption{\label{avgLSD}LSD profiles of $\iota$ Her. (a) Grand average profile. (b)-(f) Mean profiles from individual nights. Shows in each panel are Stokes $I$ (bottom of each panel, in black), Stokes $V$ (top of each panel, in red) and diagnostic null $N$ (middle of each panel, in blue). Note that $V$ and $N$ are shifted and scaled for display purposes, and the scaling differs for each panel. }
\end{centering}
\end{figure*} 

\subsection{Least Squares Deconvolution}

The diagnosis of the magnetic field is based on averaging of Zeeman signal potentially present in many spectral lines using the Least-Squares Deconvolution procedure \citep{1997MNRAS.291..658D}. In this work, we employ the iLSD ('improved LSD') procedure \citep{2010A&A...524A...5K} as implemented in a {\sc fortran 90} code written by O. Kochukhov (Uppsala), using an IDL front-end developed by J. Grunhut (ESO). The iLSD procedure provides for multi-profile deconvolution (using multiple line masks) and regularization of the deconvolution process \citep{2010A&A...524A...5K}. However, neither of these features was used in our study.

We developed several line masks based on a Vienna Atomic Line Database (VALD) \citep{2000BaltA...9..590K} {\sc extract stellar} request corresponding to the atmospheric parameters summarised in Table 1, using a line-depth cutoff of 0.01. We used an IDL tool developed by J. Grunhut to clean the line mask, removing hydrogen and He lines with broad wings, lines blended with H/He lines, and lines which had no observed counterpart in the spectrum. Of the latter there were many; a large number of predicted lines appear to be in emission, rather than absorption, or entirely absent from the spectrum. We  interactively adjusted the depths of the remaining lines to match the observed depths. We experimented with the line-depth cutoff, ultimately adopting a cutoff of 0.1. Our final mask contained approximately 130 spectral lines mainly consisting of singly-ionised lines of light elements (C, N, O, Ne, Mg, Si, etc.). 

Using this line mask, we computed LSD profiles in a window spanning $-300, 300$~km/s on a grid spacing of 1.8~km/s. We computed profiles for each individual spectrum, for the combined spectra corresponding to each night of observation, and from the combination of all spectra. All LSD profiles were normalised to the following values of the Land\'e factor, wavelength and line depth: $\bar g=1.20$, $\bar\lambda=500$~nm and $\bar d=0.1$. The typical noise level in individual LSD Stokes $V$ profiles is $7.2\times 10^{-5}~I_c$, while that in the nightly-averaged profiles ranges from $1.3\times 10^{-5}~I_c$ to $3.0\times 10^{-5}~I_c$. The noise level in the grand average profile (computed from the average of all 128 individual spectra) is $5.6\times 10^{-6}~I_c$.

The grand average and nightly averaged LSD profiles of $\iota$~Her are illustrated in Fig.~\ref{avgLSD}.

\section{Magnetic field diagnosis}

We employed the $\chi^2$ detection approach described by \citet{1992A&A...265..669D} and the detection thresholds introduced by  \citet{1997MNRAS.291..658D} to evaluate the presence of any significant signal in the LSD profiles. No significant or marginally significant signal (i.e. no false alarm probability smaller then $10^{-3}$) was obtained within the bounds of the mean spectral line for any of the individual or averaged $V$ or $N$ profiles. We therefore tentatively conclude that no significant magnetic signal is detected in our observations of $\iota$~Her.

In order to quantify our upper limits, we measured the mean longitudinal magnetic field from our LSD profiles using the first moment method, applying Eq. (1) of \citet{2000MNRAS.313..851W}. In particular, we measured each individual LSD profile, as well as the nightly means and the grand average profile. The individual profiles yield typical longitudinal field error bars of $\sim 4$~G. The nightly mean profiles have error bars ranging from 0.5 to 1.3 G. The grand average profile is characterised by an error bar of 0.3~G. None of the measurements corresponds to a longitudinal field significant at greater than 3$\sigma$. 

We also computed weighted averages of the individual profile longitudinal field measurements and the nightly mean profile measurements. We find that these weighted averages, as well as the grand average profile measurement, yield essentially identical longitudinal field values and uncertainties. This suggests that line profile variations have little impact on our ability to detect and measure the magnetic field.

The results of the $\chi^2$ detection analysis and the mean longitudinal field measurements are reported in Table~\ref{tab:log} and Table~\ref{results}.





\begin{table*}
\caption{\label{results}Longitudinal field and detection probability measurements of LSD profiles. Columns from left to right correspond to: LSD profile name, starting and ending velocities used for determination of longitudinal field and detection probability, mean error bar (per 1.8~km/s pixel) in Stokes $V$ LSD profile, detection probability, longitudinal field and longitudinal field significance in Stokes $V$, detection probability, longitudinal field and longitudinal field significance in diagnostic null $N$. The longitudinal field significance is defined as $\bz/\sigma$, where $\sigma$ is the longitudinal field formal error.} 
\begin{center}
\begin{tabular}{ccrcrrrrrrrrrr}
\hline
 &	start	&	end	&$\sigma$ & $P_V$	& $\bz$			&	$z_{\rm V}$	& $P_N$ &	$\nz$	&	$z_{\rm N}$	\\
&	\multicolumn{2}{c}{(km/s)}		& ($\times 10^{-5}$)	& (\%) &		(G)		&		& (\%)	& (G)			&		\\
\hline	
Night 1&	$-33$	&	$-6$	& 1.3&34.3&$	0.08	\pm	0.67	$&	0.1&24.8&$	-0.53	\pm	0.68	$&	$-0.8$	\\
Night 2&	$-33$	&	$-8$	& 1,2&7.9&$	0.18	\pm	0.55	$& 0.3	&19.5&$	-0.66	\pm	0.56	$&$-1.2$	\\
Night 3&	$-33$	&	$-7$	& 1.2&37.5&$	-0.61\pm	0.58	$&	$-1.1$	&3.0&$	-0.10	\pm	0.58	$&	$-0.2$	\\
Night 4&	$-33$	&	$-8$	& 3.0&7.1&$	-0.02	\pm	1.34	$&	$-0.0$	&5.1&$	-1.19	\pm	1.38	$&	$-0.1$	\\
Night 5&	$-36$	&	$-4$	& 1.6&36.1&$	-1.87	\pm	1.04	$&	$-1.8$	&2.5&$	1.41\pm	1.04	$&	1.4	\\

	\hline													
Mean ind \bz	&  \multicolumn{2}{c}{(As nightly)}	 & &&$	-0.29	\pm	0.34	$&	$-0.9$	&&$	-0.24	\pm	0.34	$&	$-0.7$	\\
Mean  	&	&		& &&$	-0.03	\pm	0.32	$&	$-0.1$	&&$	-0.11	\pm	0.32	$&$-0.3$	\\
nightly \bz\\
Grand average &	$-34$	&	$-5$	& 0.55 &42.9&$	-0.24	\pm	0.32	$&	$-0.8$	&6.7&$	-0.22	\pm	0.32	$&	$-0.7$	\\
spectrum\\
\hline
\end{tabular}
\end{center}
\end{table*}

\section{Modeling}

\subsection{Organised magnetic field}

We began by comparing the LSD profiles of each star to a grid of synthetic Stokes $V$ profiles using the method of \citet{2012MNRAS.420..773P}.
In this approach, we assume a simple centred dipolar field model, parametrised by the dipole field strength $B_d$, the rotation axis inclination $i$ with respect to the line of sight, the positive magnetic axis obliquity $\beta$, and the set of observed rotational phases $\Phi=[\varphi_1..\varphi_N]$ associated with a set of Stokes $V$ observations of a same star.

We began by fitting the shape of the Stokes $I$ LSD profile. The model \citep{2012MNRAS.420..773P} makes assumptions similar to those described in Sect. 4.4, except that due to the different codes employed, a Milne-Eddington atmosphere is used (instead of a linear limb-darkening law), and a Voigt profile (rather than a Gaussian profile) is here assumed for the shape of the local line profile. The differences in the line profile calculation resulting from these choices are entirely negligible. 

A projected rotational velocity of 6~km/s was adopted.

Given that the rotation periods are generally unknown, the rotational phases are treated as nuisance parameters by marginalizing the posterior probability density over the possible phase values in a Bayesian statistical framework. We therefore obtain the goodness-of-fit of a given rotation-independent $\mathcal{B}$=[$B_d$, $i$, $\beta$] magnetic configuration.
However, as the observations here were obtained during 4 consecutive nights, this last assumption may not valid if the rotational period is much longer than the observational timescale (i.e. if $i\gtrsim 45\degr$ as described in Sect. 2).
Therefore, we also considered the case that all the observations were obtained at the same rotational phase -- meaning that the Stokes V signature of a potential magnetic field would have been the identical for all observations -- by analysing the grand average profile.
The analysis was performed on both Stokes $V$ profiles and the null profiles.

We assume a prior probability with a $\sin(i)$ dependence for the inclination angle consistent with a random orientation of rotational axes. The obliquity angle and the rotational phases have a conservative constant prior, whereas the prior probability for the dipolar strength is chosen to be scale independent. A detailed description is provided by \citet{2012MNRAS.420..773P}.

\begin{figure}
\begin{centering}
\includegraphics[width=8cm]{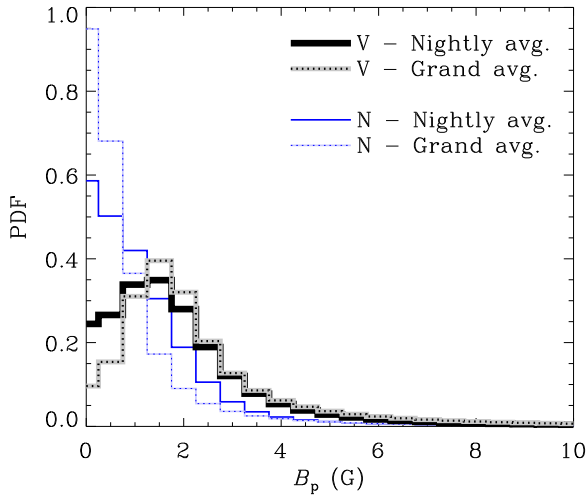}
\caption{\label{dipole}Posterior probability density functions of the magnetic dipole model marginalised for dipolar strength.}
\end{centering}
\end{figure}

Fig.~\ref{dipole} shows the resulting posterior probability density functions (PDFs) that have been marginalised for the dipolar field strength.
The nightly averages and the grand average yielded similar field strength upper limits (tabulated in Table~\ref{dipole_tab}). The upper bound of the 95.4 percent credible region is $\sim$8\,G. According to the numerical investigations of \citet{2012MNRAS.420..773P}, this threshold corresponds typically to a definite detection of a Stokes $V$ Zeeman signature.

Unlike the PDFs for the null profiles (thin blue histograms), the PDFs for the Stokes $V$ profiles (thick histograms) does not peak at 0\,G. 
However, the better fit to the observations achieved by the use of the dipole model is not statistically justified, as shown by the odds ratios (Table~\ref{dipole_tab}) which describe the ratio of the global likelihood of the non-magnetic model ($M_0$) to the magnetic dipole model ($M_1$).

We therefore conclude that the data provide no convincing evidence for the presence of a dipole magnetic field in the photosphere of $\iota$~Her, with an upper limit of ~8~G.

\begin{table}
\caption{Results from the magnetic dipole analysis. The odds ratios of the non-magnetic model ($M_0$) to the magnetic model ($M_1$) are given for the nightly observations and the grand average, for the Stokes $V$ and null $N$ profiles. The upper bound of the credible regions (in G) enclosing a certain percent of the probability density is also given.}
\label{dipole_tab}
\begin{tabular}{lrr}
\hline
                       & Nightly & Grand avg. \\
                       \hline
   $\log(M_0/M_1)$ $V$        & 0.62    & 0.31 \\
   \multicolumn{3}{c}{Credible regions $V$    (G)} \\
   68.3    \%                    &  2.3  & 2.9 \\
   95.4    \%                    &  6.7  & 8.6 \\
   99.0    \%                    & 14.9  & 17.5 \\
   99.7    \%                    & 22.5   & 28.5 \\
   \hline
   $\log(M_0/M_1)$ $N$        & 0.80    & 0.90 \\
   \multicolumn{3}{c}{Credible regions $N$    (G)} \\
   68.3    \%                    &  1.5      & 1.0 \\
   95.4    \%                    &  3.9      & 4.2 \\
   99.0    \%                    & 8.5       & 12.4 \\
   99.7    \%                    &  14.8   & 35.3 \\
   \hline
\end{tabular}
\end{table}

\subsection{Random spot fields}

\citet{2013A&A...554A..93K} examined the detectability of small-scale magnetic spots in stellar photospheres, using a model of randomly-distributed circular spots with specified filling factor, size and magnetic field strength. Fig. 5 (left panels) of their paper shows that for a $T_{\rm eft}=20000$~K star, $2\degr$ radius magnetic spots of strength 0.5~kG with a filling factor of 0.5 produce a maximum unsigned longitudinal field of approximately 10~G and a maximum amplitude of Stokes $V$ of $4\times 10^{-3}$ for $v\sin i=10$~km/s. Our grand average profile is characterised by a longitudinal field upper limit of $3\sigma_{\rm B}=1.3$~G and a Stokes $V$ upper limit of $3\sigma_{\rm V}=1.7\times 10^{-5}$, i.e. respectively 7-8 times and 235 times smaller than in the simulation. Assuming linear scaling of the longitudinal field and Stokes $V$ amplitude with spot field strength, we infer that $2\degr$ spots with strengths of 2-3~G would be detectable in our data. This analysis ignores the small differences in the LSD profile wavelength and Land\'e factor (similar in both studies) and any differences in the relative depths of the LSD Stokes $I$ profiles \citep[both are approximately 10\% deep; ][]{2013A&A...554A..93K}.

Focusing on their Fig. 7, we see that \citet{2013A&A...554A..93K} find that 2 degree spots with strengths somewhat larger than 200~G would be detectable in a typical  observation acquired in the Magnetism in Massive Stars (MiMeS ) survey \citep[][i.e. LSD SNR$\simeq $23,000]{2013arXiv1310.3965W} of a 20000~K star with $v\sin i=10$~km/s. We can estimate the sensitivity of our observations based on a direct comparison of SNRs. The SNR of our LSD profiles is $1.8\times 10^6$, i.e. nearly 80 times better than that of a typical MiMeS observation. Scaling the results using the ratio of SNRs, we infer that 2 degree spots with strengths of 3-4~G and a filling factor of 50\% should be detectable in our data. This result is consistent with that derived above.

The projected rotational velocity of $\iota$~Her is somewhat smaller than the model assumed here. However, considering that the sensitivity curves in Fig. 7 of their paper converge at small spot sizes for $v\sin i\ltsim 50$~km/s, we do not foresee that this will provide a significant detectability advantage.

\subsection{Modeling Vega-like fields}

In order to assess the feasibility of detecting a magnetic field in $\iota$ Her similar to that found in Vega, we used the ZDI magnetic field strength and geometry derived by \citet{2010A&A...523A..41P} for Vega, and generated a set of synthetic line profiles for $\iota$ Her.  These synthetic line profiles were then compared to our observed LSD profiles for $\iota$ Her.  

\citet{2010A&A...523A..41P} derived a magnetic field map for Vega using the Zeeman Doppler Imaging (ZDI) program of \citet{2006MNRAS.370..629D}.  ZDI inverts the rotationally modulated variability of a Stokes $V$ line profile, in order to reconstruct the vector magnetic field at the surface of a star.  We can take the resulting ZDI magnetic field map and apply it to a different model star, to predict what Stokes $V$ signature that star would have if it had a magnetic field identical to Vega's.

\subsubsection{Calculating synthetic $V$ profiles}

In order to interpret the magnetic map of \citet{2010A&A...523A..41P} and generate a set of synthetic Stokes $V$ profiles, we developed our own line modeling code.  This program uses the same physical assumptions as Donati et al. (2006), effectively performing the forward modeling part of their full ZDI implementation.  

This program functions by generating local line profiles for a large number of stellar surface elements, then summing these line profiles together to produce a disk integrated line profile.  Individual local line profiles in Stokes $I$ are calculated as simple Gaussians.  While this is clearly a rough approximation, rotational broadening typically dominates the shape a stellar line profile, thus this approximation has been found to be sufficient for ZDI \citep[e.g.][]{1997A&A...326.1135D} as long as the $V$ line profile signal is not very large relative to the noise.  This is particularly true when ZDI is applied to LSD profiles, which themselves are only approximations of a real line profile.  While our observed $I$ spectra have very high S/N, we detect no signal in our $V$ profiles. Therefore an approximate line model is sufficient.  Local Stokes $V$ line profiles, for individual surface elements, are calculated using the weak field approximation for Zeeman splitting:

\begin{equation}
V(\lambda) = -  \frac{e}{(4\pi m_{e} c)} \lambda_{0}^{2} g_{\rm eff} B \cos(\theta) \frac{\mathrm{d}I}{\mathrm{d}\lambda} ,
\end{equation}
where $B$ is the strength of the local magnetic field vector and $\theta$ is its angle with respect to the line of sight, $\lambda_{0}$ is the effective wavelength of the line, $g_{\rm eff}$ its effective Land\'e factor, and $I$ is the Stokes $I$ profile.  

The local line profiles are calculated for a grid of surface elements of approximately equal surface area.  The star is assumed to have a perfectly spherical geometry.  The grid is generated with the number of points in longitude ($N_\phi$) being $N_\phi = 2 N_\theta \sin \theta$, where $\theta$ is the colatitude and $N_\theta$ is the number of points in colatitude. The local profiles are then Doppler shifted according to the projection of their rotational velocity along the line of sight.  The star is assumed to be in solid body rotation.  The Doppler shifted profiles are then summed, weighted by the area of the surface element projected on the sky, and by the element's relative brightness.  A linear limb darkening law \citep[e.g.][]{2005oasp.book.....G} is used to compute the brightness of the surface elements.  The $V$ and $I$ profiles are then divided by the disk integrated continuum flux to produce normalised $V/I_c$ and $I/I_c$ profiles.  Finally, the disk integrated line profiles are convolved with a Gaussian instrumental profile to match the observed spectral resolution.  

The magnetic vector for each surface element on the star was calculated from a table of spherical harmonic coefficients, as described by Donati et al. (2006, eqs 2-4).  This set of coefficients comprise the ZDI magnetic field map, and are the principal output of the ZDI code used to map Vega.  These magnetic coefficients are interpreted in an identical fashion to that used by \citet{2010A&A...523A..41P} and Donati et al. (2006).  
For clarity, we note that there are two significant differences between the equations 2-8 as stated by Donati et al. (2006) and the actual implementation in their code (for a star rotating in the right handed sense).  Equation 2 is actually implemented without the leading minus sign (i.e. $B_{r} = \sum_{l,m} \alpha_{l,m} Y_{l,m}(\theta, \phi)$).  Additionally, the actual implementation of the (complex valued) coefficients $\alpha_{l,m}$, $\beta_{l,m}$, and $\gamma_{l,m}$ uses the complex conjugate of the values stated by Donati et al. (2006). 
As a further minor clarification, the Legendre polynomials $P_{l,m}(x)$ (following the common definition of e.g. \citet{1965hmfw.book.....A} are actually functions of $\cos \theta$, where $\theta$ is the stellar colatitude, rather than just $\theta$ as written in Donati et al. (2006). 

These differences have no impact on the accuracy of the ZDI code, or the maps it produces.  They only change the sign of the (typically unpublished) coefficients.  Thus, for the large majority of applications this difference is trivial, however the correct interpretation of a specific set of $\alpha_{l,m}$, $\beta_{l,m}$, and $\gamma_{l,m}$ is important for accurate recovery of the surface field distribution.  

The accuracy of the synthetic line profiles generated by this code has been checked extensively against the code of Donati et al. (2006).  Various magnetic maps computed from real Stokes $V$ time series using the code of Donati et al.\ were used as input for our code (largely maps from solar twins investigated by Folsom et al.\ in prep.), and our synthetic line profiles were compared to those generated by the  Donati et al.\ ZDI code as part of the inversion process.  Identical model line  parameters (Gaussian width, depth, wavelength, and effective Land\'e factor) and model stellar parameters (rotational phase, $i$, $v \sin i$, and limb darkening) were used, and identical model line profiles were successfully produced.  
These tests were run for stars with a variety of magnetic field geometries, inclination angles, and projected rotational velocities, and perfect agreement was always found.  

\subsubsection{Model line profiles for Vega}

To verify our methodology, we generated model line profiles for Vega, using the magnetic geometry and stellar parameters described by \citet{2010A&A...523A..41P}.  
{In this study we focus on the July 2008 map they presented, since there is no clear evidence for the evolution of the magnetic field between their 2008 and 2009 maps.  The observations in June 2008 from \citet{2010A&A...523A..41P} detect a Stokes $V$ signature at a higher level of significance than the Sept.\ 2009 observations, thus we prefer the magnetic map based on these data.  Additionally, the magnetic geometry in the 2008 map is simpler, and thus the map is more likely correct from a maximum entropy standpoint.  }
As a consequence of the significant $v \sin i$ of Vega (22 \kms\ in our model), we reconstruct substantial line profile variability over a rotational cycle, despite the very small inclination of the rotation axis to the line of sight ($i = 7^\circ$).  
We computed a rotationally averaged Stokes $V$ profile by taking the mean of 50 line profiles evenly spaced in rotational phase. 
The shapes of our mean line profiles agree very well with those shown in Fig.\ 7 of \citet{2010A&A...523A..41P}.  The amplitude of our mean line profile also agrees with Petit et al. (2010), once we used the magnetic map with the correct amplitudes from Petit et al. (2014, {\bf erratum}), which contain a peak magnetic field strength of $\sim$7.8 G in 2008. 

The smaller scale, more rapidly varying features in the synthetic line profile, corresponding to smaller scale features further from the rotational axis in the magnetic map, are less significant.  These features are generally not detected in individual LSD profiles, and thus it is hard to confirm that these features are not simply noise.  However, the strong, nearly constant $V$ signature near the center of the line, corresponding to the strong polar radial magnetic spot in the map, appears clearly in the mean LSD profile.  Therefore, we have the most confidence in the accuracy of this feature.  

We consider two scenarios for comparison with our observations: one in which we take the full Vega map from 2008, and one in which we retain only the most reliable features of the map. In this second case, we consider only the components of the map that are symmetric about the stellar rotation axis, since these are the components that contribute significantly to the rotation averaged line profile, and are thus clearly seen in the mean LSD profile.  The non-axisymmetric components of the map largely cancel out over a rotation cycle, and thus are not detected with a very high confidence.  We also ignore the toroidal axisymmetric components of the map, since at the inclination of Vega's rotation axis, these are virtually always perpendicular to the line of sight, and hence unconstrained by Stokes $V$ observations.  In practice the modified map was created by keeping only the $\alpha$ and $\beta$ coefficients with $m=0$ in the map (see Donati et al.\ 2006 eq.\ 2-4) and setting all other coefficients to zero. 
We find that a synthetic line profile generated using only the axisymmetric poloidal magnetic field components reproduces the rotationally averaged line profile with a high degree of accuracy.  
The magnetic geometries used in these two scenarios are illustrated in Fig 4, for two different inclinations.

\begin{figure*}
\centering
\includegraphics[width=1.7in]{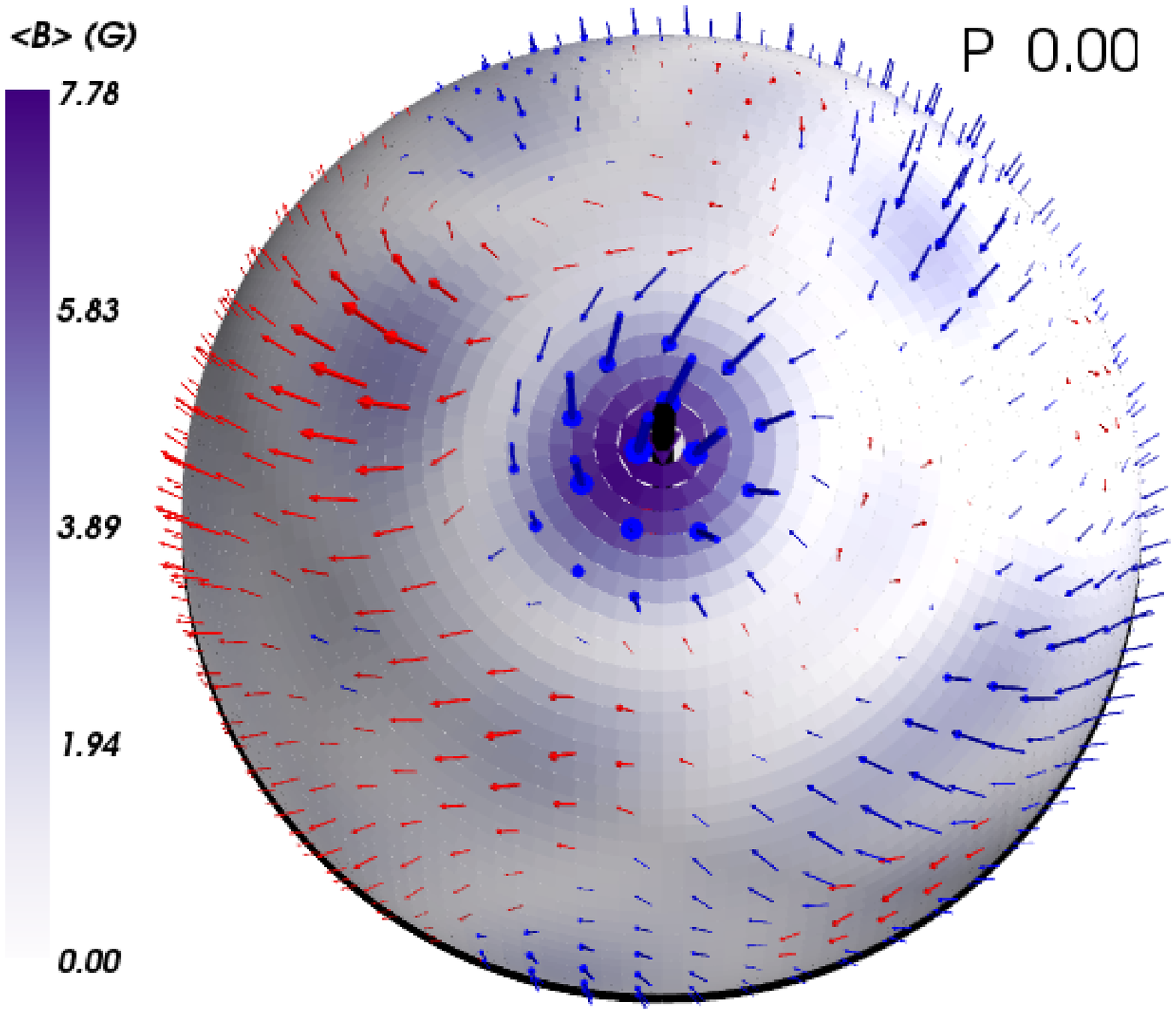}
\includegraphics[width=1.7in]{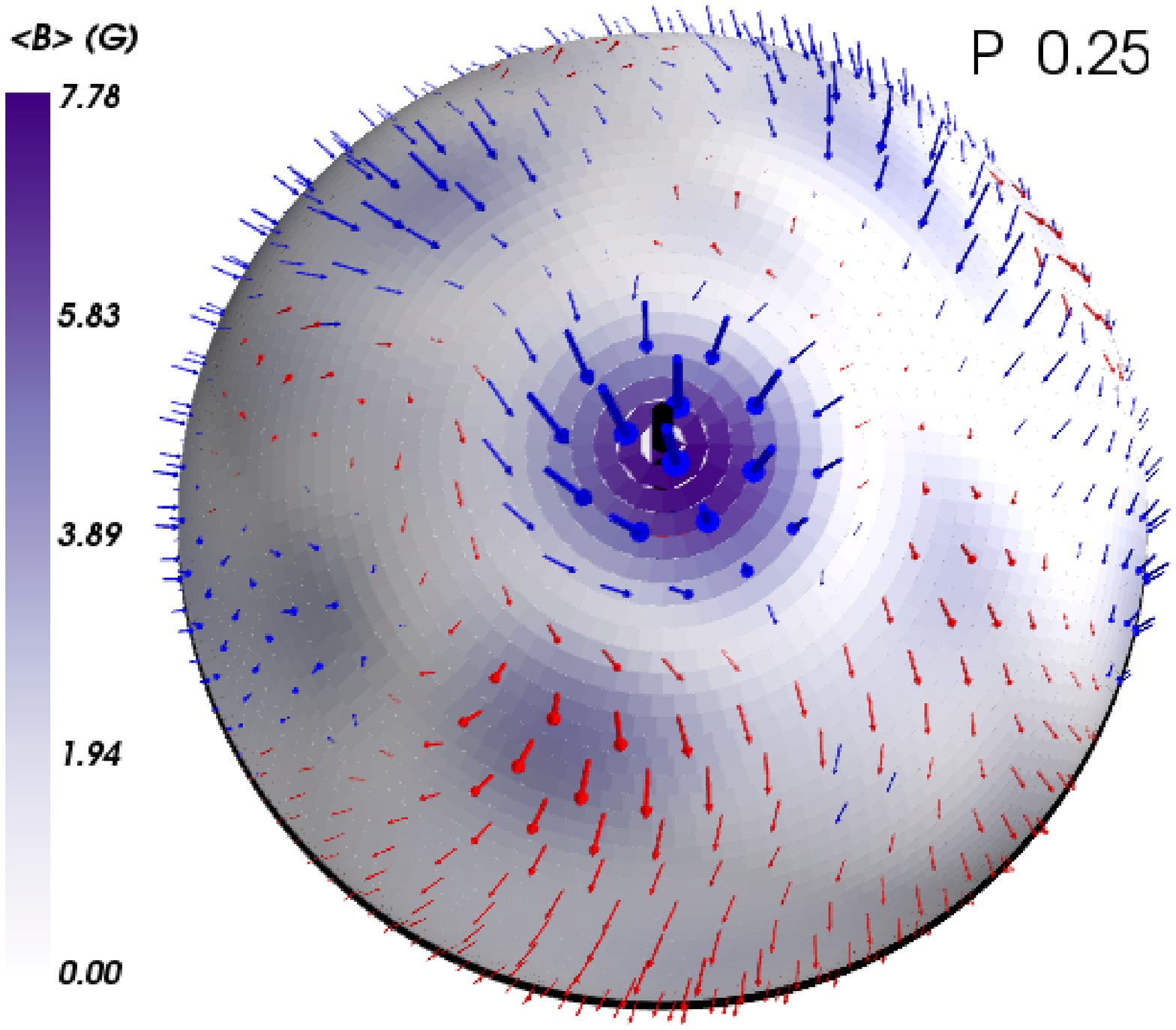}
\includegraphics[width=1.7in]{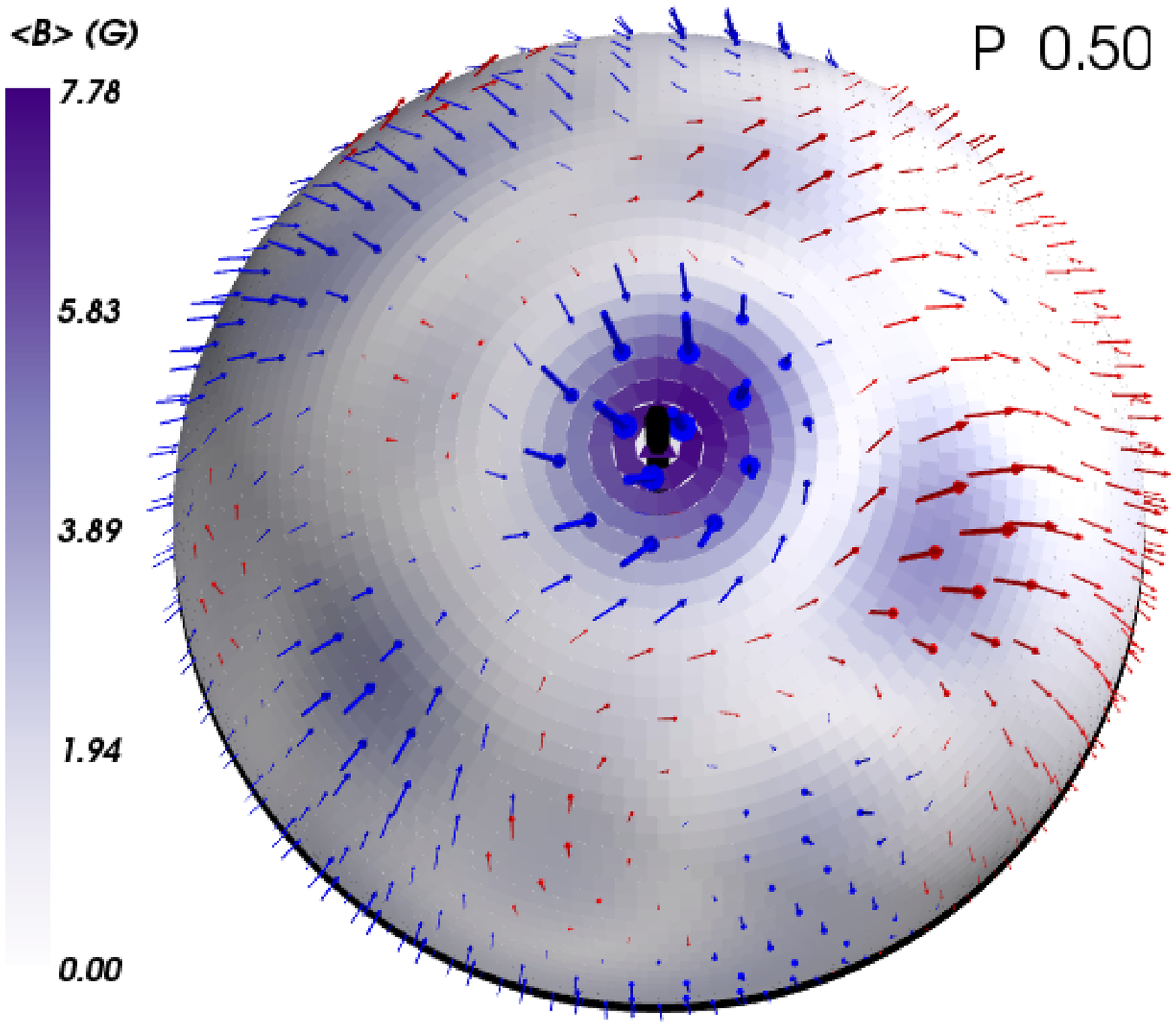}
\includegraphics[width=1.7in]{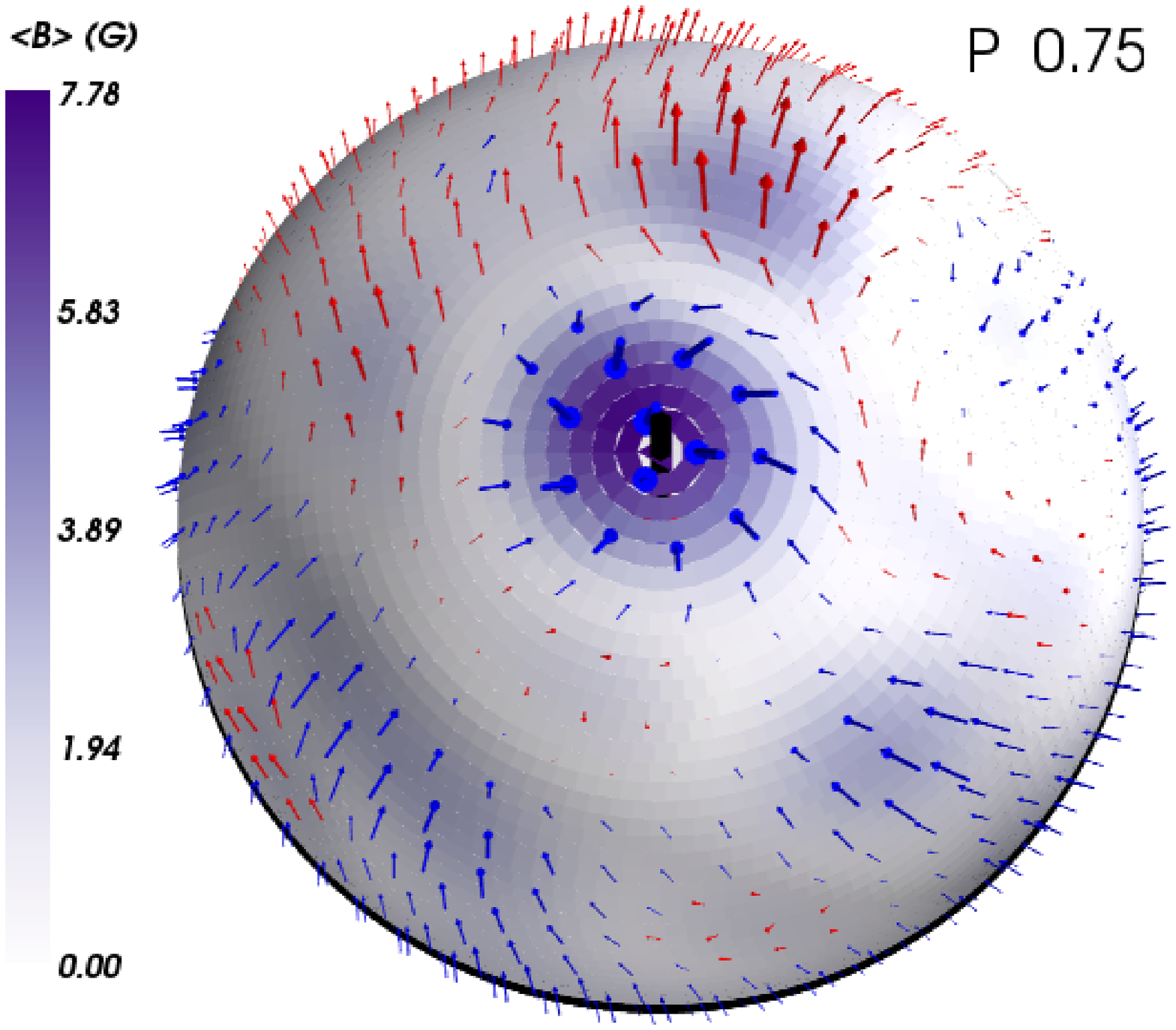}
\includegraphics[width=1.7in]{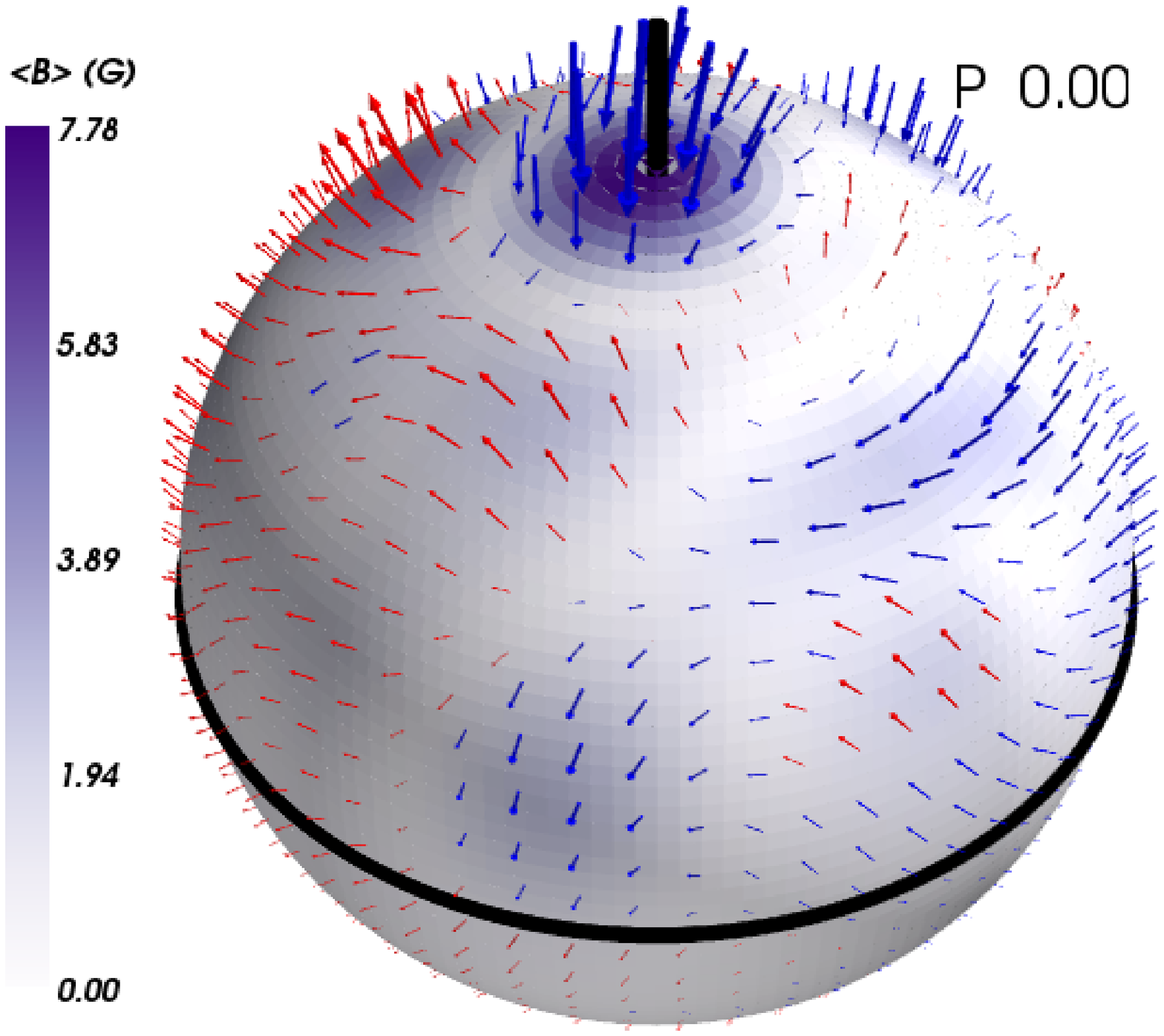}
\includegraphics[width=1.7in]{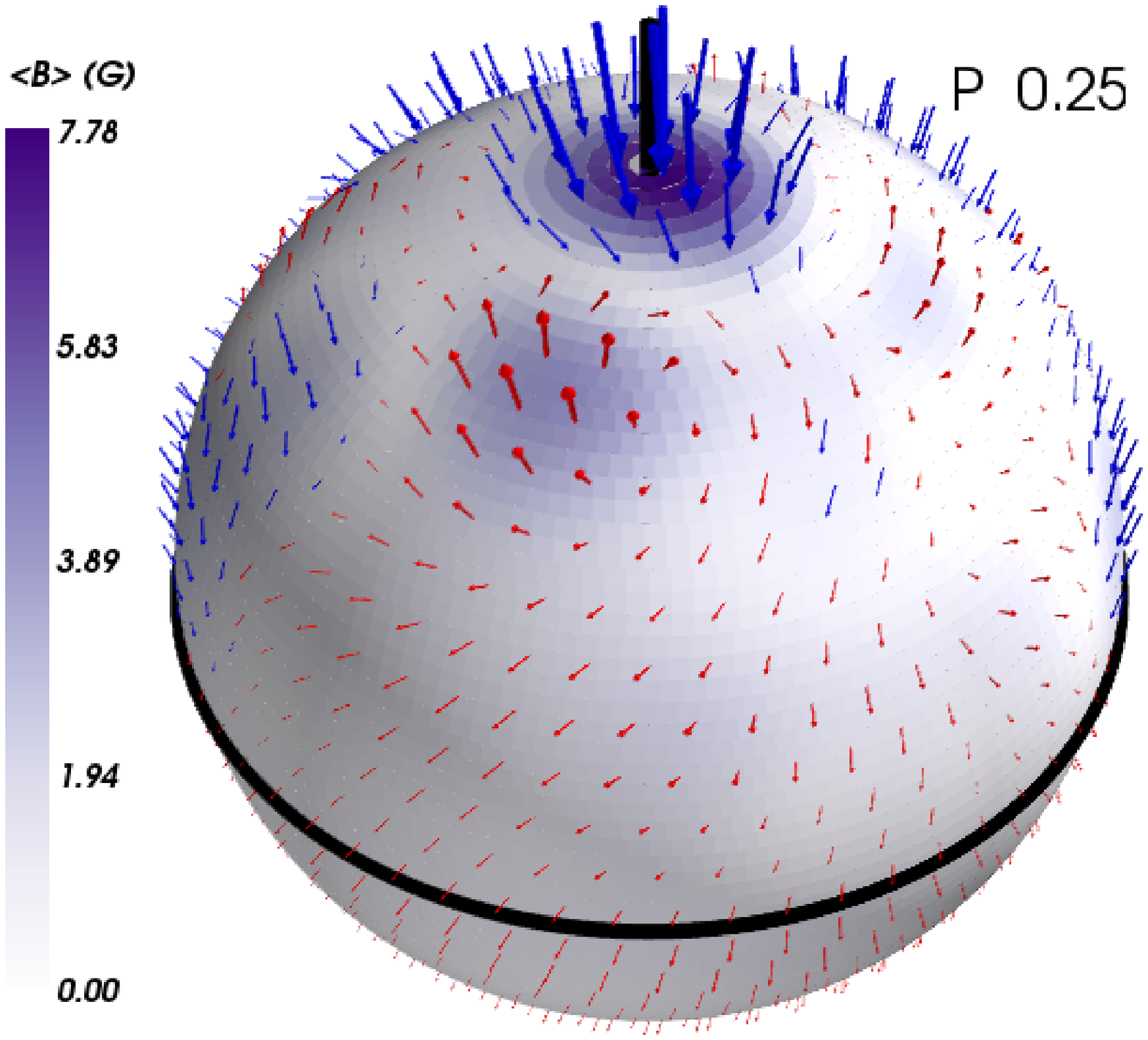}
\includegraphics[width=1.7in]{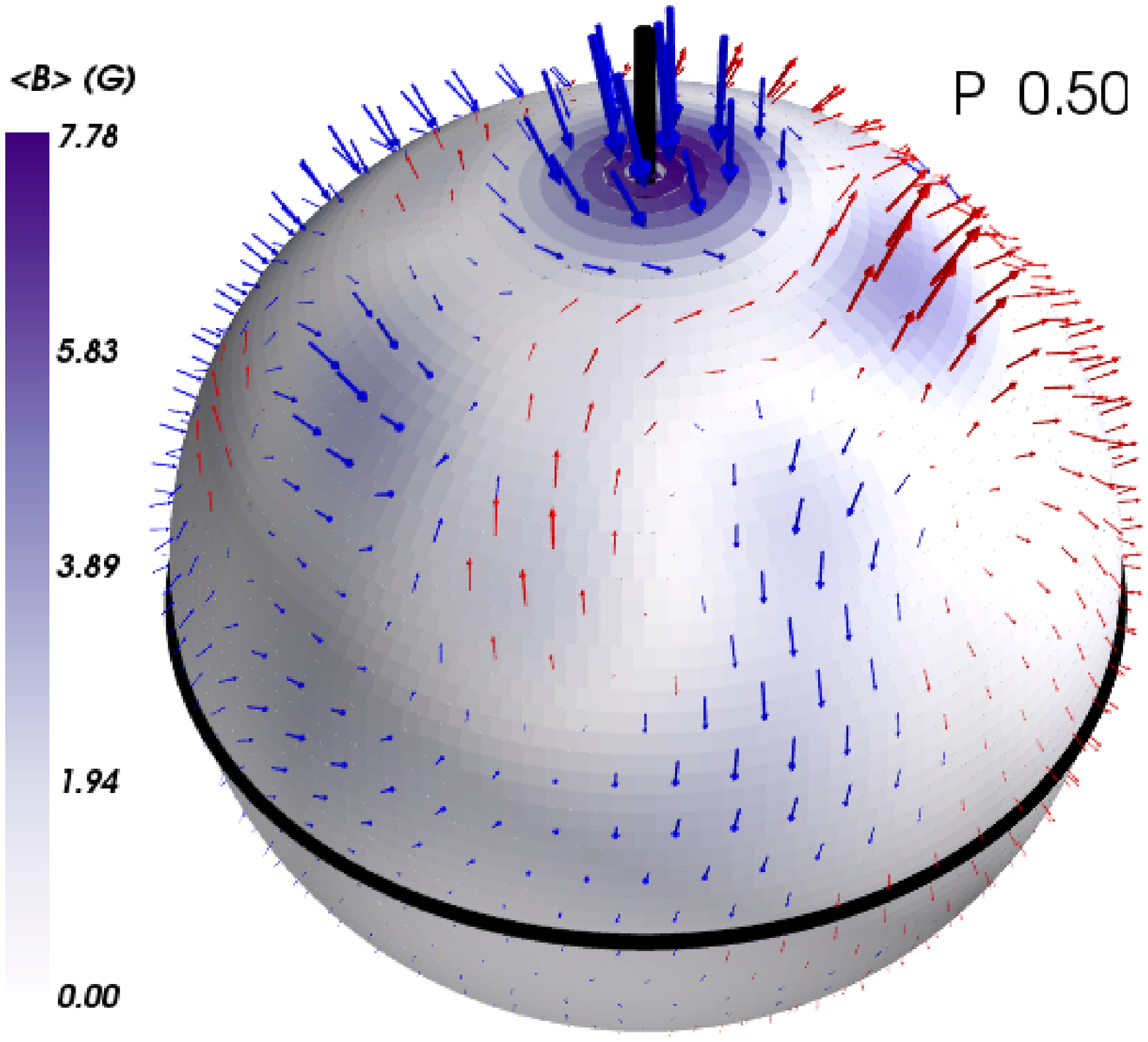}
\includegraphics[width=1.7in]{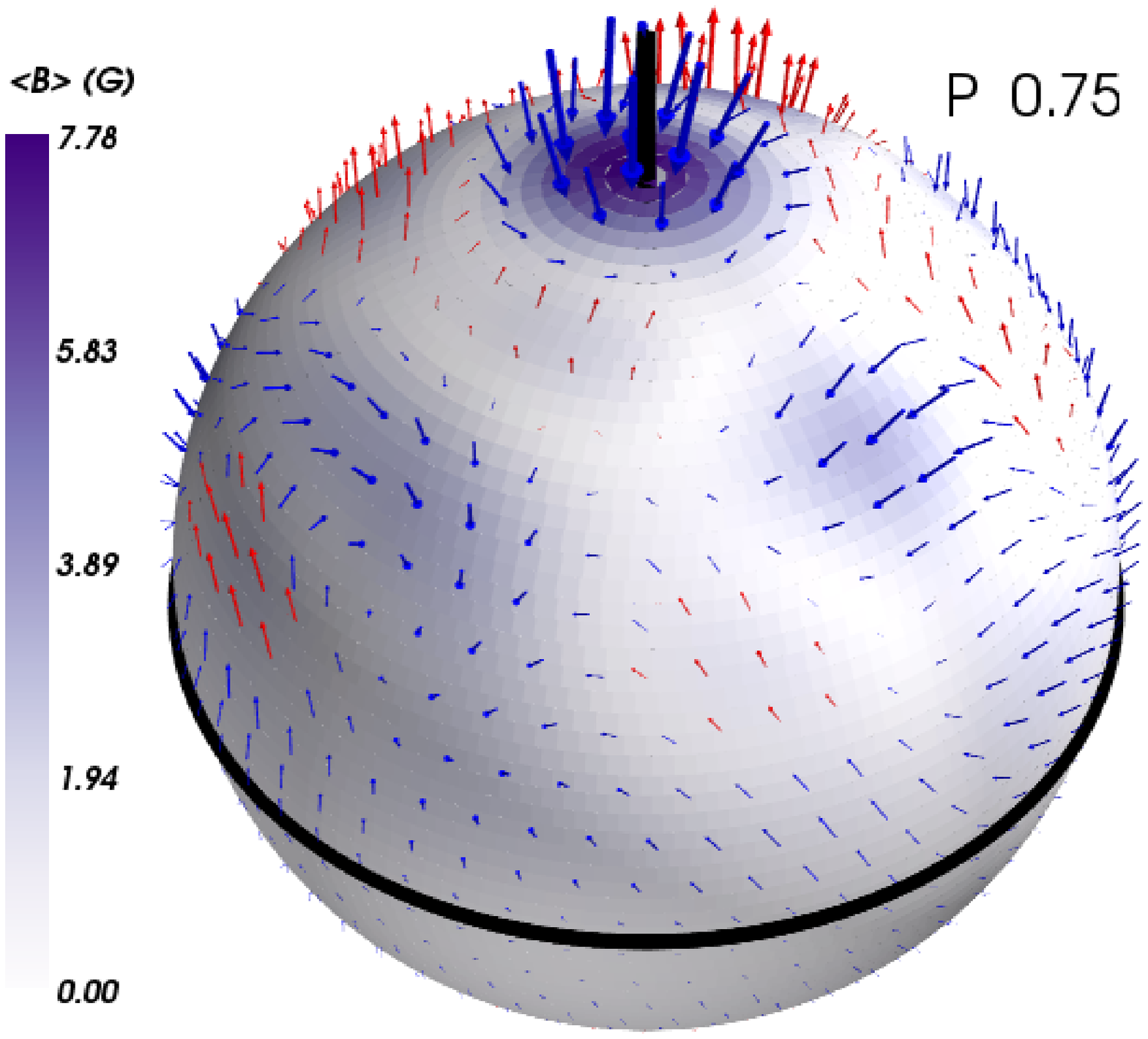}
\includegraphics[width=1.7in]{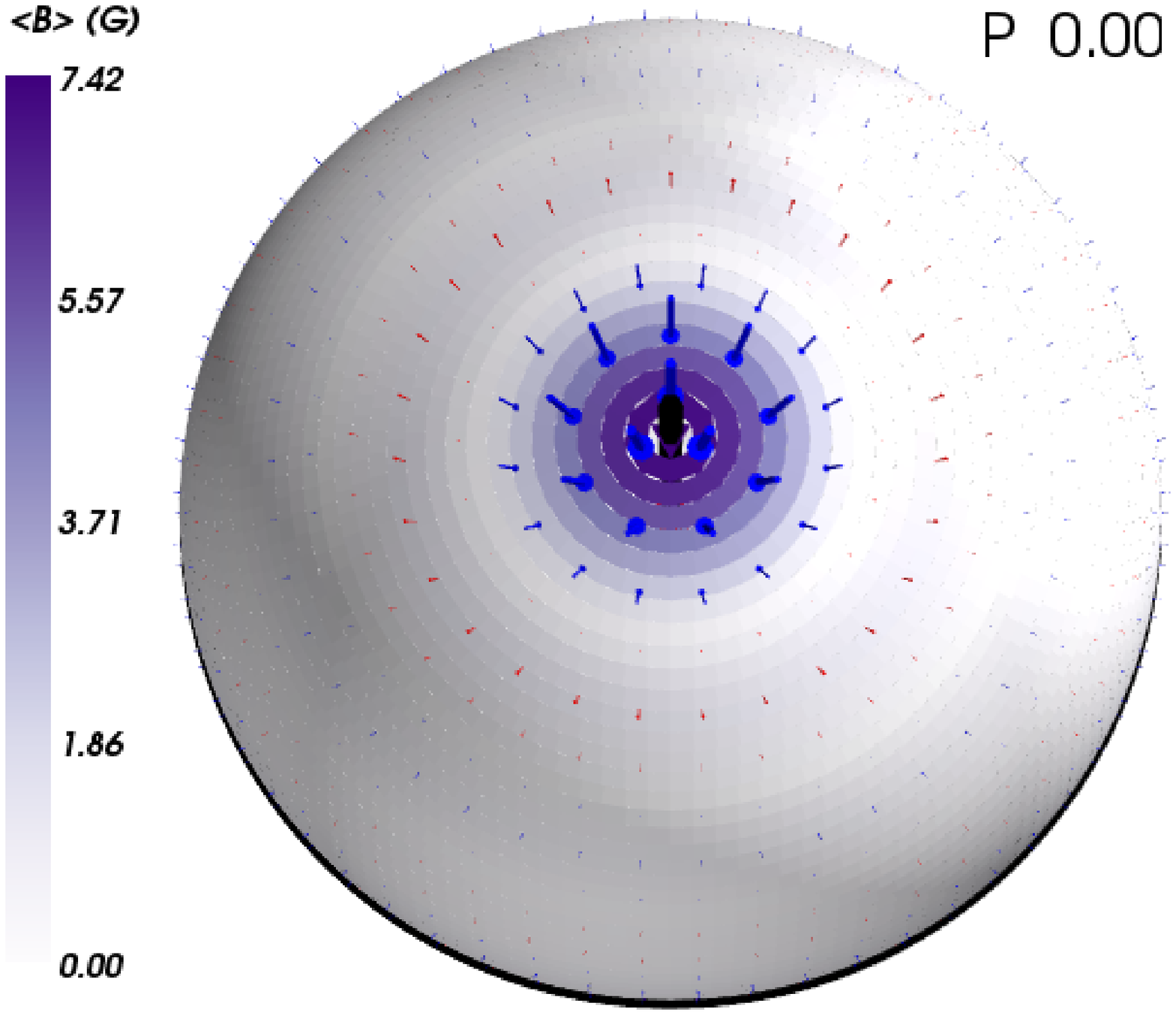}
\includegraphics[width=1.7in]{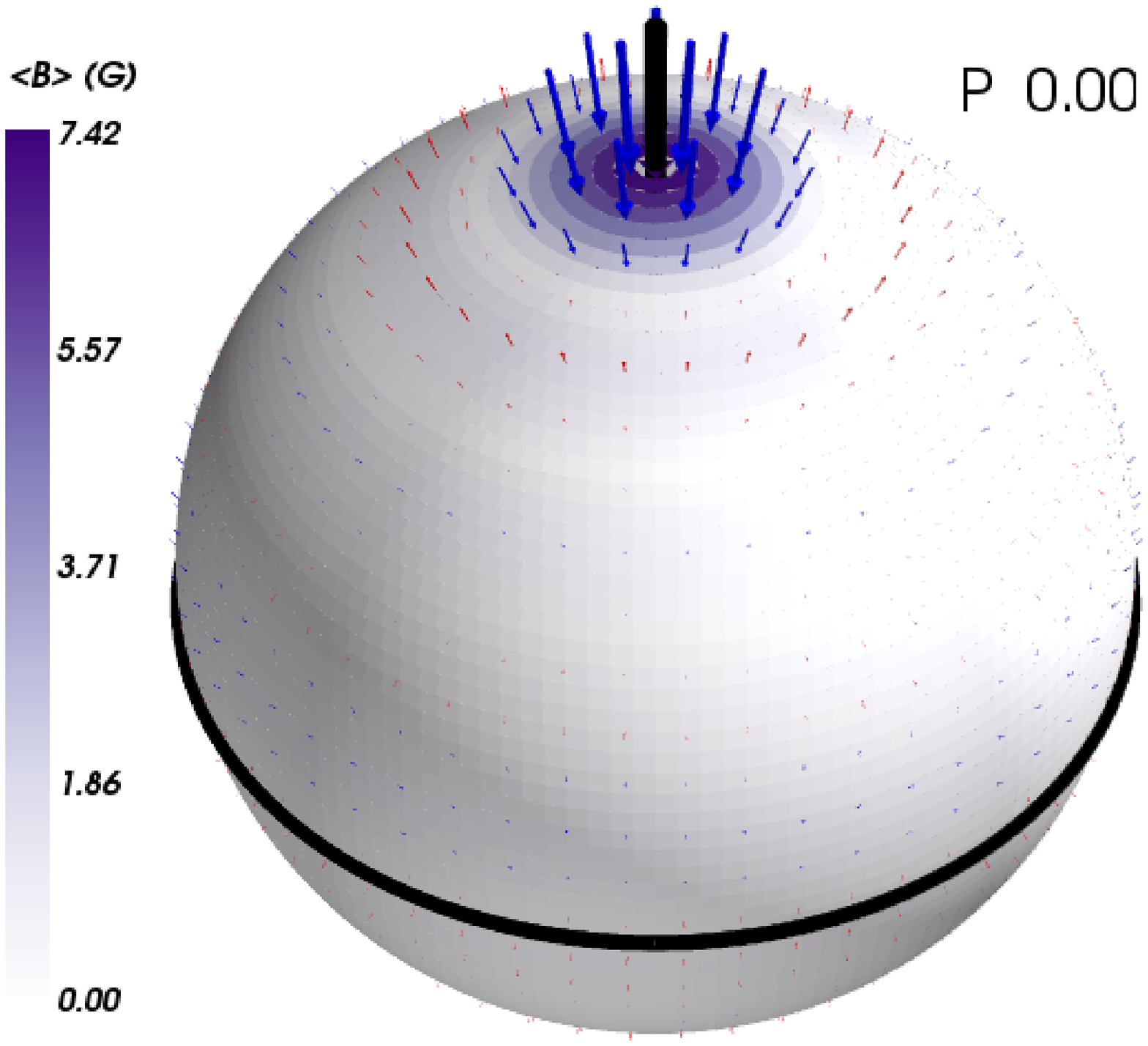}
\caption{Illustrations of the magnetic field geometries used. The top row is the 2008 Vega magnetic map at $i=7^\circ$, the second row is the same map at $i=45^\circ$.  The last row presents the axisymmetric poloidal components of the 2008 map at $i=7^\circ$ and $i=45^\circ$. Black lines indicate the rotation pole and equator.  Arrows represent the direction of the magnetic field, with blue arrows having a negative radial component and red arrows having a positive radial component.  Frames are labeled by rotational phase.  }
\label{fig-mag-geoms}
\end{figure*}

\subsubsection{Model line profiles for $\iota$ Her}

\begin{figure*}
\centering
\includegraphics[width=3.4in]{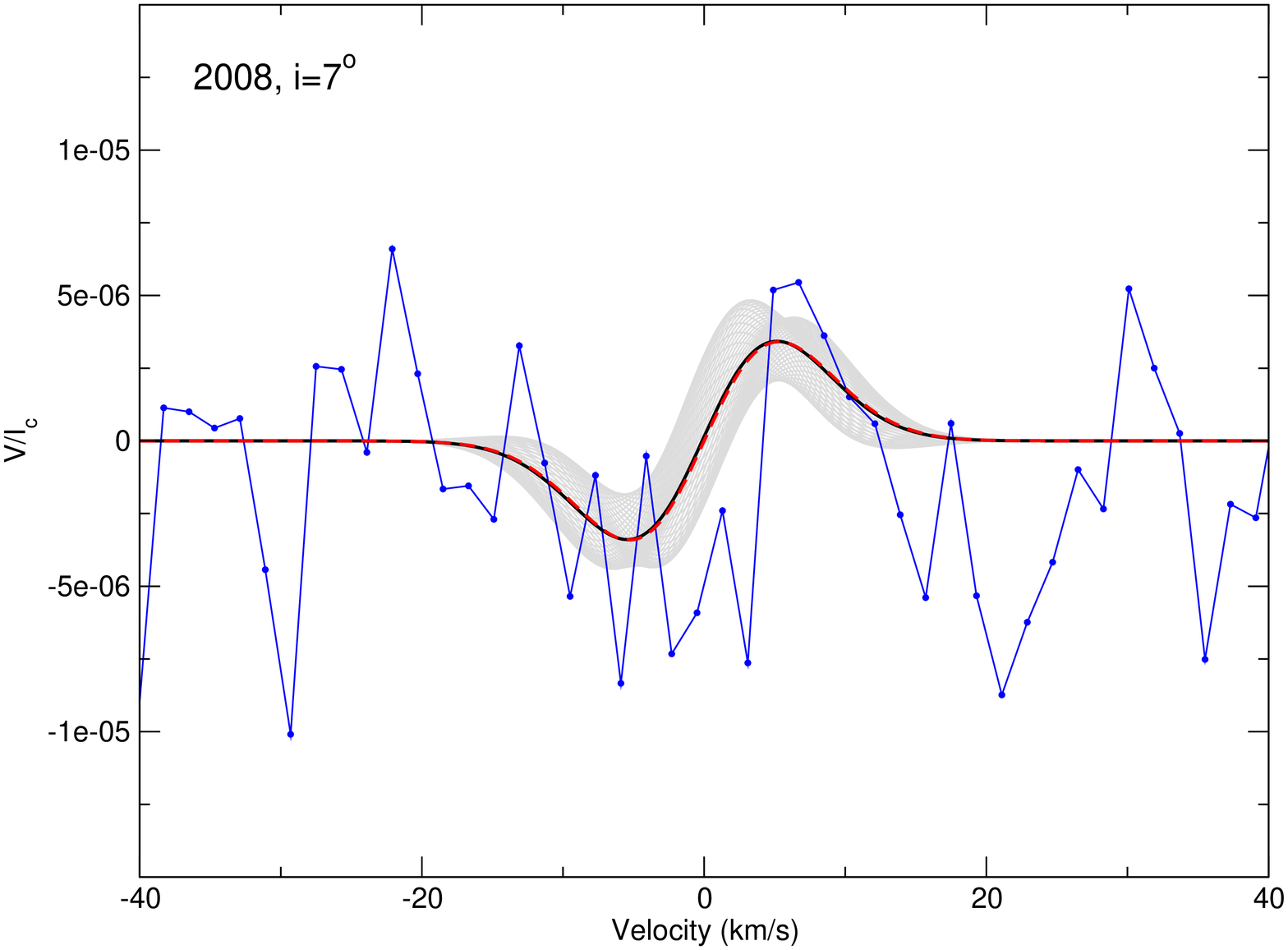}
\includegraphics[width=3.4in]{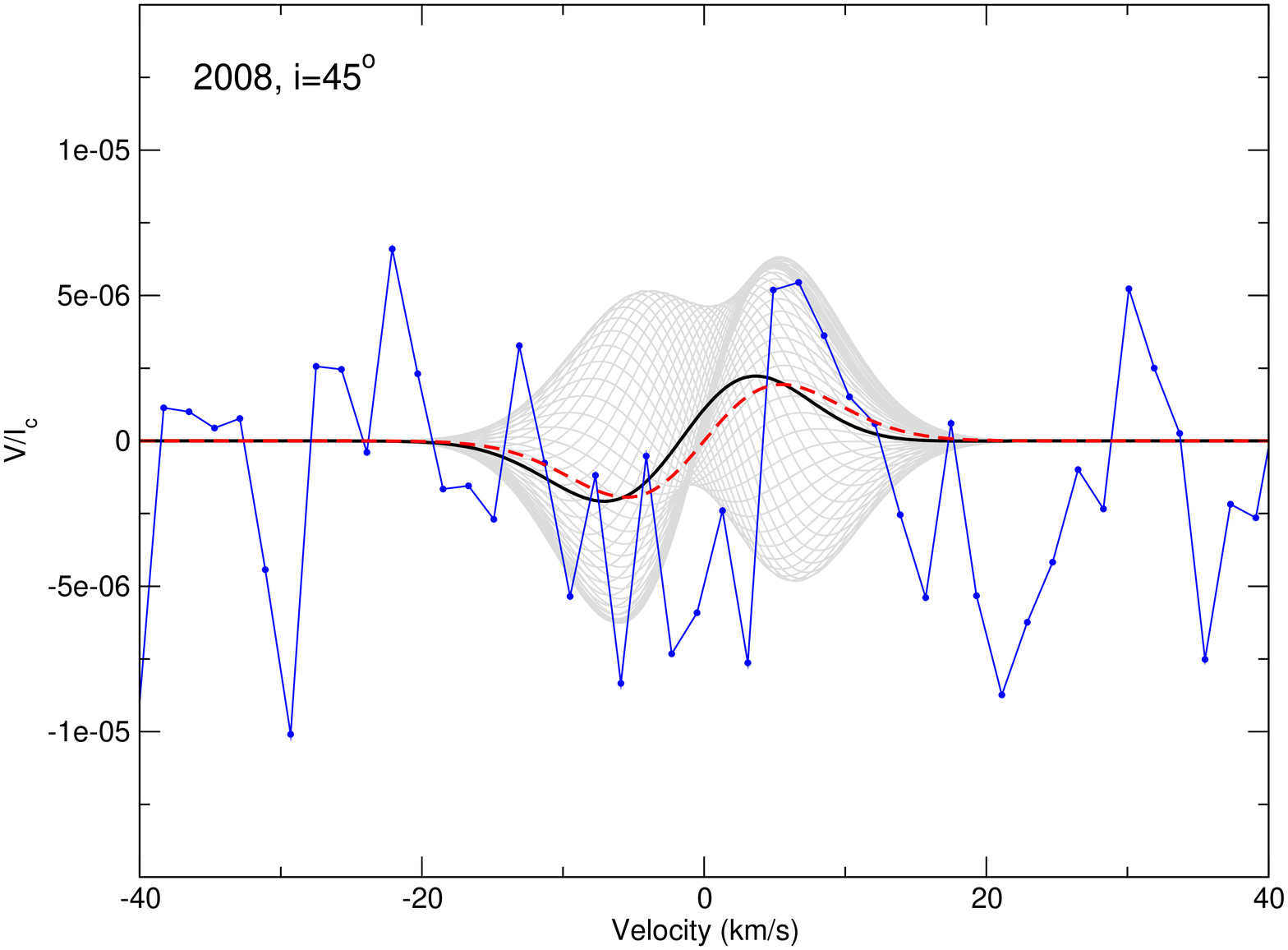}
\caption{Model Stokes $V$ line profiles for $\iota$ Her, based on the magnetic field map of Vega in 2008 from Petit et al. (2010).  Grey lines are models for individual rotation phases, thick black lines are the line profile averaged over the rotation cycle.  Thick dashed (red) lines are models computed using only the axisymmetric poloidal components of the magnetic field map.  The thin line with points (blue) is our observed mean $V$ LSD profile.  Models are computed for inclinations of the rotation axis of $7^\circ$ and $45^\circ$, as labeled. }
\label{model-lsdprofs}
\end{figure*}

$\iota$ Her has a reasonably precise $v \sin i$ of $6 \pm 1$ \kms \citep{2012A&A...539A.143N}, however the inclination of the stellar rotation axis is unknown.  Therefore we consider two scenarios, one in which the star has the same inclination as Vega ($i = 7^{\circ}$), and one in which the star has a $45^{\circ}$ inclination.   For randomly oriented rotational axes, it is much more probable to find an inclination near $45^{\circ}$ than $7^{\circ}$.  Our model line uses a centre wavelength of 500 nm and a Land\'e factor of 1.2, matching the normalization values used to calculate the LSD profiles.  We set the local line depth by fitting the LSD $I$ line profile (a depth of 0.10 of the continuum at line center). We set the local line width to 5 \kms, by fitting the line widths of several individual lines in the observed spectra, as well as fitting the width of the LSD profile.  We used a limb darkening coefficient of 0.75 (a typical value for a late-B star; Gray 2005), and stellar grid with 8146 surface elements.  $v \sin i$ was set to 6 \kms and an instrumental resolving power of 65000 was used. 

A $45^{\circ}$ inclination angle, coupled with the known $v \sin i$ implies a rotation period in the 26 - 40 day range, as discussed in Sect. 2.1.  Our observations were obtained over a period of 4 days, and thus would cover between 10\%  and 16\% of a rotation cycle if $\iota$ Her had a $45^{\circ}$ inclination.  This implies that phase smearing in our grand average LSD profile would be fairly minor at this inclination.  A $7^{\circ}$ inclination implies a rotation period in the 4.6 - 6.9 day range.  For this inclination our observations would span between 60\% and 90\% of a rotation cycle, thus our grand average LSD profile would be close to a rotationally averaged line profile in this model.   

$\iota$ Her has a significantly lower $v \sin i$ than Vega, effectively providing less resolution of the stellar surface.  As a consequence, there is significantly more cancellation in the $V$ profile from surface regions with different magnetic polarity.  This lowers the amount of rotational variability in the profile and produces a less complex line profile, but the amplitude of the $V$ profile does not change strongly as a consequence of $v \sin i$.  

Increasing the inclination of the star effectively reduces the accuracy of the magnetic model.  The magnetic map of Vega is essentially only constrained on one hemisphere, with the magnetic field in regions below the equator being driven to zero by the regularization chosen.  Regions near the equator, but still in the visible hemisphere of Vega, are weakly constrained due to increasing limb darkening and a radial field component that is nearly perpendicular to the line of sight.  Thus, as the inclination increases, more of the stellar surface with weak, unconstrained magnetic field values becomes visible.  Furthermore, as the strong polar spot becomes more inclined with respect to the line of sight, weaker, less well constrained magnetic features in the Vega map contribute proportionally more to the model $V$ profile.   For randomly oriented rotation axes, the mean inclination angle would be $\sim$60$^{\circ}$.  However, as a consequence of the inaccuracy of our map at such large inclinations, we do not consider any inclinations beyond $45^{\circ}$.  

{In order to assess the impact of systematic uncertainties in the magnetic map on our model line profiles, we computed model line profiles using the 2009 Vega magnetic field map of \citet{2010A&A...523A..41P}.  For $i = 7^{\circ}$ the rotationally averaged 2009 line profiles were qualitatively similar to the 2008 line profiles, although with slightly lower amplitude.  However, at $i = 45^{\circ}$ the rotationally averaged line profile for the 2009 map had a much lower amplitude than the 2008 map, while the line profiles at individual phases had significantly larger amplitudes in the 2009 map and somewhat different shapes.  This supports the conclusion that systematic uncertainties are modest in our $i = 7^{\circ}$ models, but large enough to be significant in our $i = 45^{\circ}$ models.  }

In Fig.~\ref{model-lsdprofs} we show synthetic $V$ profiles for the Vega map at $i = 7^{\circ}$ and $i = 45^{\circ}$, both for many individual rotational phases and for rotationally averaged profiles.  For the $i = 7^{\circ}$ models we find an amplitude that is comparable to the noise in our mean observed LSD $V$ profile.  In the $i = 45^{\circ}$ case we find slightly higher amplitudes at some phases, but the rotationally averaged profile has a slightly lower amplitude, and again the profiles have amplitudes near that of the noise in the observation.  However, since this inclination increases the importance of less reliable features in the map, the large amplitudes at individual phases may not be accurate. 
Using a magnetic map based on the axisymmetric poloidal components only, we find $V$ profiles very similar to the rotational average of the full Vega map at both inclination angles.  The inclination angle chosen implies a degree of phase smearing in our observed mean LSD profile.  For the $i = 7^{\circ}$ case, where our observations would span nearly a full rotation cycle, the rotationally averaged model is the most realistic.  For the $i = 45^{\circ}$ case, where our observations would cover only 10-15\% of the rotation cycle, the models at individual phases are more realistic. In the case of a $45\degr$ inclination, there are in fact some phases at which a magnetic field similar to Vega's would have been detected in our observations.

\begin{figure}
\centering
\includegraphics[width=3.2in]{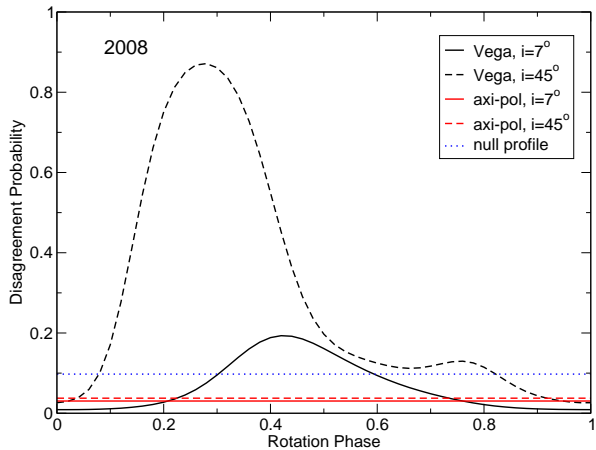}
\caption{Probabilities that model Stokes $V$ line profiles are inconsistent with the observations, as a function of the rotational phase of the computed profile.  These are presented for the 2008 Vega map, for inclinations of the rotation axis of $7^\circ$ and $45^\circ$.  The `axi-pol' models are based on only the axisymmetric poloidal components of the Vega magnetic map, and the `null profile' model corresponds to no signal in Stokes $V$. }
\label{fig-prob-detect-phase}
\end{figure}

To evaluate the detectability of a magnetic field identical to Vega's in a more quantitative fashion, we calculated reduced $\chi^2$ values for the observed LSD profile relative to the models.  Based on $\chi^2$ statistics we evaluated the probability that our model profiles are inconsistent with the observations.  This provides an approximation of the probability that we would have detected the signal in a model if it were in our observation.  These probabilities were calculated for models using both inclinations, for a range of rotational phases, presented in Fig. \ref{fig-prob-detect-phase}.  The models based on the axisymmetric components of the map are included as well, but they do not vary with rotation.  The `null profile' model is provided for reference, and it is a model with no signal in Stokes $V$.  We also calculated disagreement probabilities for the rotationally averaged line profiles at both inclinations.  
In the rotationally averaged cases, we find no significant probability of `detecting' our model line ($i = 7^\circ$, $P = 0.033$; $i = 45^\circ$, $P = 0.072$), with values similar to the corresponding axisymmetric models.  
At some rotational phases for the $i=45^\circ$ model the disagreement probability becomes larger, but it never exceeds 90\%.

\begin{figure}
\centering
\includegraphics[width=3.2in]{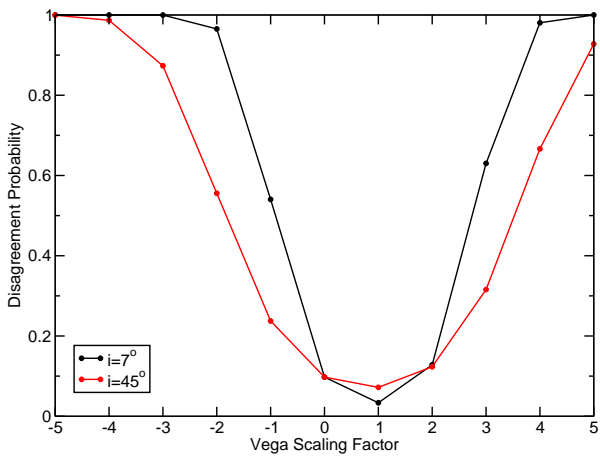}
\caption{Probabilities that model Stokes $V$ line profiles are inconsistent with the observations, using rotationally averaged model line profiles, based on the Vega magnetic field geometry, scaled by various factors.  A $7^\circ$ inclination (black) and $45^\circ$ inclination (red/gray) are shown. }
\label{fig-prob-detect-scaling}
\end{figure}

From this modeling, we conclude that it is unlikely we would have detected a magnetic field identical to that of Vega in $\iota$ Her, unless we observed at a particularly favourable inclination and rotation phase.  However, we are near the limit of detection for such a field.  We therefore consider a few models in which we scale up the strength of the magnetic field in the Vega map but leave the morphology unchanged.  In these tests we consider the 2008 magnetic map of \citet{2010A&A...523A..41P} at inclinations of $7^\circ$ and $45^\circ$, and for simplicity we only use the rotationally averaged line profile.  We consider a scaling of up to five times the strength of Vega, we also consider scaling the magnetic field with an opposite sign.  
At $7^\circ$ the rotationally averaged line profile is a good approximation, and with this we find a scaling factor $\geq 4$ or $\leq -2$ provides models inconsistent with our observations at a high ($> 90\%$) confidence.  
At  $45^\circ$, again using the averaged line profile, a scaling factor $> 5$ or $\leq -4$ produces models inconsistent with our observations (at $> 90\%$).  For $i=45^\circ$ our observations have some phase resolution, thus the averaged line profile is not as good an approximation.  However, for this inclination, nearly all individual phases produce line profiles with larger amplitudes and larger disagreement probabilities than the rotationally averaged profile (see Fig.~\ref{model-lsdprofs} and Fig.~\ref{fig-prob-detect-phase}).  Thus for the $i=45^\circ$ case, these scaling values are effectively upper limits on the scaling needed to produce a detectable profile.  We conclude that while it is unlikely we would have detected a magnetic field identical to Vega's, we would have likely detected one with a peak strength of about 30 G, i.e. approximately four times as strong as that of Vega.

\section{Summary, discussion and conclusions}

In this investigation we have analyzed 128 high resolution and signal-to-noise ratio Stokes $V$ spectra of the sharp-lined B3IV star $\iota$~Her, obtained on 5 consecutive nights in 2012. Using Least-Squares Deconvolution with a carefully selected line mask, we computed LSD Stokes $I$, Stokes $V$ and $N$ profiles from individual spectra, nightly-averaged spectra, and the grand mean spectrum. In no case was any evidence of a magnetic field detected. We obtained a formal uncertainty associated with the longitudinal magnetic field of the grand mean of $\pm 0.3$~G.

We modelled the LSD profiles in 3 different ways. First, we used the Bayesian approach of \citet{2012MNRAS.420..773P}, inferring an upper limit on any dipole field present of 8~G. Secondly, we compared our LSD profiles with the calculations of \citet{2013A&A...554A..93K} for random magnetic spots, inferring an upper limit of 2-3~G on magnetic spots with angular radii of $2\degr$ and a filling factor of 50\%. Finally, we compared our profiles with those predicted by the magnetic field model of Vega derived by \citet{2010A&A...523A..41P}. Realisations of the predicted Stokes $V$ profiles corresponding to their model were computed for two assumed rotational axis inclinations of $\iota$~Her: 7$\degr$ and 45$\degr$. We concluded that it is unlikely we would have detected a magnetic field identical to Vega's. However, a field four times as strong as Vega's would likely have been identified.

An important uncertainty that is only approximately addressed by our model is the rotational axis inclination of $\iota$~Her. The inclination not only defines the observational geometry of the star; it also implies (in combination with the known projected rotational velocity and stellar radius) the rotational period of the star. Therefore knowledge of the inclination is a critical parameter required to know the expected amplitudes and modulation of Stokes $V$ signatures produced by a given field configuration, as well as the degree of rotational averaging of those profiles in observations such as ours. Currently, the inclination of $\iota$~Her is unknown. However, detailed asteroseismic studies of this SPB variable star could lead to important constraints on this valuable parameter.

Our analysis of the nightly mean and grand average LSD profiles yields an upper limit of 8~G on any dipole magnetic field present in the photosphere of $\iota$~Her. In the context of the 'magnetic dichotomy' \citep{2014arXiv1402.5362L}, our null result for $\iota$~Her provides one of the tightest constraints yet on the absence of organised magnetic fields in non-Ap/Bp stars in this part of the HR diagram  \citep[compare with, e.g.][]{2002A&A...392..637S, 2010A&A...523A..40A,2014arXiv1402.5362L}.

{ Recently, \citet{2014A&A...562A..59N} published very high S/N spectropolarimetric observations of the B2IV star $\gamma$ Peg.  $\gamma$ Peg displays both $\beta$ Cep and SPB pulsation modes (Handler et al. 2009), and it appears to be an intrinsically slow rotator ($v_{\rm eq} \sim 3$ \kms; Handler et al. 2009).  In their observations with the Narval instrument at the T\'elescope Bernard Lyot, Neiner et al. (2014) found no evidence of a magnetic field in the star, with an average longitudinal magnetic field of $-0.1 \pm 0.4$ G, and concluded that it is unlikely that $\gamma$ Peg hosts a Vega-like magnetic field.  With the $V$ profile modeling methodology described above, we can more rigorously examine the possibility of a Vega-like magnetic field in $\gamma$ Peg, by using the magnetic map for Vega from Petit et al. (2010) to generate model line profiles for $\gamma$ Peg.  

To generate line models for $\gamma$ Peg, we began with the 2008 Vega magnetic map from Petit et al. (2010), which is likely the more reliable map.  For the model line, we used a wavelength of 497 nm and Land\'e factor of 1.22, matching the normalization values of the mean LSD profile from the Narval observations (Neiner, priv. comm.).  The line depth was set by fitting the depth of the mean Narval $I$ LSD profile.  We used the same local line width as for $\iota$ Her (5 \kms), and we used the same limb darkening coefficient (0.75).  We used $v \sin i = 0$ \kms\ (Telting et al. 2006), although tests with $v \sin i = 3$ \kms\ make no significant difference.  We adopt R = 65000, matching that of Narval, and used a stellar grid with 8146 surface elements.  The inclination of $\gamma$ Peg is unknown, thus we consider both the $i=7^\circ$ and $i=45^\circ$ cases.  The Narval observations of Neiner et al. (2014) span one month and likely cover the majority of a rotation cycle, possibly multiple cycles, thus we only consider rotationally averaged line profiles (created by averaging line profiles from 50 evenly spaced phases).  

Our rotationally averaged $V$ line profile for $\gamma$ Peg has an amplitude of $6\times10^{-6}$ of the continuum level for $i=7^\circ$, and $4\times10^{-6}$ for $i=45^\circ$.  In their mean LSD profile of $\gamma$ Peg from their Narval observations, Neiner et al. (2014) find a noise level of approximately $1\times10^{-5}$ of the continuum.  Therefore we conclude that, if $\gamma$ Peg had a magnetic field identical to that of Vega, it is unlikely Neiner et al. (2014) would have detected it.  However, their observations are close to the detection threshold for the Vega magnetic field model, much like our observations for $\iota$ Her.  Thus, while it is unlikely Neiner et al. (2014) would have detected a magnetic field identical to Vega's in $\gamma$ Peg, if the field were a few times stronger they likely would have detected it.  }

\citet{2009A&A...499..279C} and \citet{2011A&A...534A.140C} proposed that convection due to the iron opacity bump at subphotospheric temperatures ($\sim 10^5$~K) could produce magnetic spots in the photospheres of hot stars. However, according to Fig. 3 of \citet{2011A&A...534A.140C}, stars with HR diagram positions similar to $\iota$~Her are predicted to host magnetic spots with field strength of order 1~G. Given that the scale of such spots is not well established, it is unclear if our upper limits on random spot distributions provide meaningful constraints on these predictions.


In the context of their "failed fossil" model, \citet{2013MNRAS.428.2789B} make rather clear predictions regarding Vega-like fields: that all intermediate-mass and high-mass stars should host surface fields of at least the strength found in Vega; that those fields should decrease in strength slowly over the course of the main sequence lifetime; that faster rotators should have stronger fields; and that the length scale of the magnetic structures on the surface may be small in very young stars but should quickly increase to at least a fifth of the stellar radius. It is not entirely clear to what extent our results for $\iota$~Her are in conflict with this model. This star is not located near the beginning of the main sequence, hence the magnetic structures on its surface should be rather large (and hence more easily detectable). On the other hand, its slow rotation and relatively late main sequence location (and hence larger radius: about 70\% larger than on the ZAMS) should produce a relatively weak field.

Ultimately, we conclude that no evidence of magnetic field is detected in $\iota$~Her at the achieved precision. This result (and that of \citet{2014A&A...562A..59N} for $\gamma$ Peg) demonstrate that deep circular polarization spectra of early-type stars acquired with ESPaDOnS and Narval and analyzed using the LSD method are able to achieve null results. In itself, this provides additional confidence that positive detections at similar effective SNRs (like those obtained for Vega and Sirius) are not instrumental artefacts. 

Investigations of this kind represent an important probe of the characteristics of magnetism in stars with radiative envelopes, potentially leading to new breakthroughs in our understanding of the physics of the phenomenon of 'fossil' magnetism.

\bibliography{iher}

\begin{thebibliography}{36}
\expandafter\ifx\csname natexlab\endcsname\relax\def\natexlab#1{#1}\fi

\bibitem[{{Abramowitz} \& {Stegun}(1965)}]{1965hmfw.book.....A}
{Abramowitz} M., {Stegun} I.~A., 1965, {Handbook of mathematical functions with
  formulas, graphs, and mathematical tables}

\bibitem[{{Auri{\`e}re} {et~al}\mbox{.}(2010){Auri{\`e}re}, {Wade},
  {Ligni{\`e}res}, {Hui-Bon-Hoa}, {Landstreet}, {Iliev}, {Donati}, {Petit},
  {Roudier}, \& {Th{\'e}ado}}]{2010A&A...523A..40A}
{Auri{\`e}re} M. {et~al.}, 2010, \aap, 523, A40

\bibitem[{{Auri{\`e}re} {et~al}\mbox{.}(2007){Auri{\`e}re}, {Wade},
  {Silvester}, {Ligni{\`e}res}, {Bagnulo}, {Bale}, {Dintrans}, {Donati},
  {Folsom}, {Gruberbauer}, {Hui Bon Hoa}, {Jeffers}, {Johnson}, {Landstreet},
  {L{\`e}bre}, {Lueftinger}, {Marsden}, {Mouillet}, {Naseri}, {Paletou},
  {Petit}, {Power}, {Rincon}, {Strasser}, \& {Toqu{\'e}}}]{2007A&A...475.1053A}
{Auri{\`e}re} M. {et~al.}, 2007, \aap, 475, 1053

\bibitem[{{Babcock}(1958)}]{1958ApJS....3..141B}
{Babcock} H.~W., 1958, \apjs, 3, 141

\bibitem[{{Bagnulo} {et~al}\mbox{.}(2009){Bagnulo}, {Landolfi}, {Landstreet},
  {Landi Degl'Innocenti}, {Fossati}, \& {Sterzik}}]{2009PASP..121..993B}
{Bagnulo} S., {Landolfi} M., {Landstreet} J.~D., {Landi Degl'Innocenti} E.,
  {Fossati} L., {Sterzik} M., 2009, \pasp, 121, 993

\bibitem[{{Bailey} \& {Landstreet}(2013)}]{2013A&A...551A..30B}
{Bailey} J.~D., {Landstreet} J.~D., 2013, \aap, 551, A30

\bibitem[{{Braithwaite} \& {Cantiello}(2013)}]{2013MNRAS.428.2789B}
{Braithwaite} J., {Cantiello} M., 2013, \mnras, 428, 2789

\bibitem[{{Cantiello} \& {Braithwaite}(2011)}]{2011A&A...534A.140C}
{Cantiello} M., {Braithwaite} J., 2011, \aap, 534, A140

\bibitem[{{Cantiello} {et~al}\mbox{.}(2009){Cantiello}, {Langer}, {Brott}, {de
  Koter}, {Shore}, {Vink}, {Voegler}, {Lennon}, \&
  {Yoon}}]{2009A&A...499..279C}
{Cantiello} M. {et~al.}, 2009, \aap, 499, 279

\bibitem[{{Chapellier} {et~al}\mbox{.}(2000){Chapellier}, {Mathias}, {Le
  Contel}, {Garrido}, {Le Contel}, \& {Valtier}}]{2000A&A...362..189C}
{Chapellier} E., {Mathias} P., {Le Contel} J.-M., {Garrido} R., {Le Contel} D.,
  {Valtier} J.-C., 2000, \aap, 362, 189

\bibitem[{{Donati} \& {Brown}(1997)}]{1997A&A...326.1135D}
{Donati} J.-F., {Brown} S.~F., 1997, \aap, 326, 1135

\bibitem[{{Donati} {et~al}\mbox{.}(2006){Donati}, {Howarth}, {Jardine},
  {Petit}, {Catala}, {Landstreet}, {Bouret}, {Alecian}, {Barnes}, {Forveille},
  {Paletou}, \& {Manset}}]{2006MNRAS.370..629D}
{Donati} J.-F. {et~al.}, 2006, \mnras, 370, 629

\bibitem[{{Donati} {et~al}\mbox{.}(1997){Donati}, {Semel}, {Carter}, {Rees}, \&
  {Collier Cameron}}]{1997MNRAS.291..658D}
{Donati} J.-F., {Semel} M., {Carter} B.~D., {Rees} D.~E., {Collier Cameron} A.,
  1997, \mnras, 291, 658

\bibitem[{{Donati}, {Semel} \& {Rees}(1992){Donati}, {Semel}, \&
  {Rees}}]{1992A&A...265..669D}
{Donati} J.-F., {Semel} M., {Rees} D.~E., 1992, \aap, 265, 669

\bibitem[{{Gray}(2005)}]{2005oasp.book.....G}
{Gray} D.~F., 2005, {The Observation and Analysis of Stellar Photospheres}

\bibitem[{{Handler}(2009)}]{2009MNRAS.398.1339H}
{Handler} G., 2009, \mnras, 398, 1339

\bibitem[{{Kochukhov}, {Makaganiuk} \& {Piskunov}(2010){Kochukhov},
  {Makaganiuk}, \& {Piskunov}}]{2010A&A...524A...5K}
{Kochukhov} O., {Makaganiuk} V., {Piskunov} N., 2010, \aap, 524, A5

\bibitem[{{Kochukhov} \& {Sudnik}(2013)}]{2013A&A...554A..93K}
{Kochukhov} O., {Sudnik} N., 2013, \aap, 554, A93

\bibitem[{{Kupka} {et~al}\mbox{.}(2000){Kupka}, {Ryabchikova}, {Piskunov},
  {Stempels}, \& {Weiss}}]{2000BaltA...9..590K}
{Kupka} F.~G., {Ryabchikova} T.~A., {Piskunov} N.~E., {Stempels} H.~C., {Weiss}
  W.~W., 2000, Baltic Astronomy, 9, 590

\bibitem[{{Landstreet} {et~al}\mbox{.}(2009){Landstreet}, {Kupka}, {Ford},
  {Officer}, {Sigut}, {Silaj}, {Strasser}, \&
  {Townshend}}]{2009A&A...503..973L}
{Landstreet} J.~D., {Kupka} F., {Ford} H.~A., {Officer} T., {Sigut} T.~A.~A.,
  {Silaj} J., {Strasser} S., {Townshend} A., 2009, \aap, 503, 973

\bibitem[{{Lefever} {et~al}\mbox{.}(2010){Lefever}, {Puls}, {Morel}, {Aerts},
  {Decin}, \& {Briquet}}]{2010A&A...515A..74L}
{Lefever} K., {Puls} J., {Morel} T., {Aerts} C., {Decin} L., {Briquet} M.,
  2010, \aap, 515, A74

\bibitem[{{Ligni{\`e}res} {et~al}\mbox{.}(2014){Ligni{\`e}res}, {Petit},
  {Auri{\`e}re}, {Wade}, \& {B{\"o}hm}}]{2014arXiv1402.5362L}
{Ligni{\`e}res} F., {Petit} P., {Auri{\`e}re} M., {Wade} G.~A., {B{\"o}hm} T.,
  2014, ArXiv e-prints

\bibitem[{{Ligni{\`e}res} {et~al}\mbox{.}(2009){Ligni{\`e}res}, {Petit},
  {B{\"o}hm}, \& {Auri{\`e}re}}]{2009A&A...500L..41L}
{Ligni{\`e}res} F., {Petit} P., {B{\"o}hm} T., {Auri{\`e}re} M., 2009, \aap,
  500, L41

\bibitem[{{Mathias} \& {Waelkens}(1995)}]{1995A&A...300..200M}
{Mathias} P., {Waelkens} C., 1995, \aap, 300, 200

\bibitem[{{Neiner} {et~al}\mbox{.}(2012){Neiner}, {Landstreet}, {Alecian},
  {Owocki}, {Kochukhov}, {Bohlender}, \& {MiMeS
  Collaboration}}]{2012A&A...546A..44N}
{Neiner} C., {Landstreet} J.~D., {Alecian} E., {Owocki} S., {Kochukhov} O.,
  {Bohlender} D., {MiMeS Collaboration}, 2012, \aap, 546, A44

\bibitem[{{Neiner} {et~al}\mbox{.}(2014){Neiner}, {Monin}, {Leroy}, {Mathis},
  \& {Bohlender}}]{2014A&A...562A..59N}
{Neiner} C., {Monin} D., {Leroy} B., {Mathis} S., {Bohlender} D., 2014, \aap,
  562, A59

\bibitem[{{Nieva}(2013)}]{2013A&A...550A..26N}
{Nieva} M.-F., 2013, \aap, 550, A26

\bibitem[{{Nieva} \& {Przybilla}(2012)}]{2012A&A...539A.143N}
{Nieva} M.-F., {Przybilla} N., 2012, \aap, 539, A143

\bibitem[{{Petit} {et~al}\mbox{.}(2010){Petit}, {Ligni{\`e}res}, {Wade},
  {Auri{\`e}re}, {B{\"o}hm}, {Bagnulo}, {Dintrans}, {Fumel}, {Grunhut},
  {Lanoux}, {Morgenthaler}, \& {Van Grootel}}]{2010A&A...523A..41P}
{Petit} P. {et~al.}, 2010, \aap, 523, A41

\bibitem[{{Petit} \& {Wade}(2012)}]{2012MNRAS.420..773P}
{Petit} V., {Wade} G.~A., 2012, \mnras, 420, 773

\bibitem[{{Power} {et~al}\mbox{.}(2008){Power}, {Wade}, {Auri{\`e}re},
  {Silvester}, \& {Hanes}}]{2008CoSka..38..443P}
{Power} J., {Wade} G.~A., {Auri{\`e}re} M., {Silvester} J., {Hanes} D., 2008,
  Contributions of the Astronomical Observatory Skalnate Pleso, 38, 443

\bibitem[{{Shorlin} {et~al}\mbox{.}(2002){Shorlin}, {Wade}, {Donati},
  {Landstreet}, {Petit}, {Sigut}, \& {Strasser}}]{2002A&A...392..637S}
{Shorlin} S.~L.~S., {Wade} G.~A., {Donati} J.-F., {Landstreet} J.~D., {Petit}
  P., {Sigut} T.~A.~A., {Strasser} S., 2002, \aap, 392, 637

\bibitem[{{Silvester} {et~al}\mbox{.}(2012){Silvester}, {Wade}, {Kochukhov},
  {Bagnulo}, {Folsom}, \& {Hanes}}]{2012MNRAS.426.1003S}
{Silvester} J., {Wade} G.~A., {Kochukhov} O., {Bagnulo} S., {Folsom} C.~P.,
  {Hanes} D., 2012, \mnras, 426, 1003

\bibitem[{{Wade} {et~al}\mbox{.}(2000){Wade}, {Donati}, {Landstreet}, \&
  {Shorlin}}]{2000MNRAS.313..851W}
{Wade} G.~A., {Donati} J.-F., {Landstreet} J.~D., {Shorlin} S.~L.~S., 2000,
  \mnras, 313, 851

\bibitem[{{Wade} {et~al}\mbox{.}(2013){Wade}, {Grunhut}, {Alecian}, {Neiner},
  {Auriere}, {Bohlender}, {David-Uraz}, {Folsom}, {Henrichs}, {Kochukhov},
  {Mathis}, {Owocki}, {Petit}, \& {the MiMeS
  Collaboration}}]{2013arXiv1310.3965W}
{Wade} G.~A. {et~al.}, 2013, ArXiv e-prints

\bibitem[{{Wright}(2013)}]{2013arXiv1309.6970W}
{Wright} N.~J., 2013, ArXiv e-prints

\end{thebibliography}

\clearpage

\begin{table}
\addtolength{\tabcolsep}{-1pt}
\caption{Spectropolarimetric observations. Columns report Heliocentric Julian date, signal-to-noise ratio per 1.8~km/s pixel, measured longitudinal magnetic field from Stokes $V$ and detection significance (as described in the paper), measured longitudinal magnetic field from diagnostic null and detection significance.}
\label{tab:log}
\begin{center}
\begin{tabular}{ccrrrrrrrrrrrr}
\hline
\#	&	HJD	&	S/N	& 	$\bz$			&	$z_V$	&	$\nz$			&	$z_N$	\\
\hline	
1543202	&	2456103.8733	&	996	& $	3.7	\pm	4	$ & $	0.91	$& $	-5.4	\pm	4	$ & $	-1.34	$\\
1543206	&	2456103.8778	&	1025	& $	-3.1	\pm	4	$ & $	-0.79	$& $	-3.3	\pm	4	$ & $	-0.82	$\\
1543210	&	2456103.8824	&	988	& $	-0.2	\pm	4	$ & $	-0.05	$& $	-2.3	\pm	4	$ & $	-0.56	$\\
1543214	&	2456103.8870	&	995	& $	4.3	\pm	4	$ & $	1.05	$& $	3.4	\pm	4	$ & $	0.82	$\\
1543218	&	2456103.8915	&	950	& $	7.0	\pm	4	$ & $	1.65	$& $	-1.4	\pm	4	$ & $	-0.32	$\\
1543222	&	2456103.8960	&	1017	& $	0.0	\pm	4	$ & $	0.01	$& $	7.9	\pm	4	$ & $	1.99	$\\
1543226	&	2456103.9006	&	1054	& $	-3.1	\pm	4	$ & $	-0.81	$& $	-2.5	\pm	4	$ & $	-0.64	$\\
1543230	&	2456103.9051	&	1089	& $	-1.8	\pm	4	$ & $	-0.48	$& $	2.2	\pm	4	$ & $	0.59	$\\
1543234	&	2456103.9096	&	1079	& $	-2.9	\pm	4	$ & $	-0.77	$& $	-4.0	\pm	4	$ & $	-1.05	$\\
1543238	&	2456103.9143	&	1036	& $	-0.7	\pm	4	$ & $	-0.17	$& $	-1.7	\pm	4	$ & $	-0.44	$\\
1543242	&	2456103.9188	&	1048	& $	7.8	\pm	4	$ & $	1.99	$& $	1.4	\pm	4	$ & $	0.34	$\\
1543246	&	2456103.9233	&	974	& $	5.5	\pm	4	$ & $	1.36	$& $	-2.0	\pm	4	$ & $	-0.50	$\\
1543250	&	2456103.9279	&	1014	& $	-4.5	\pm	4	$ & $	-1.10	$& $	-0.4	\pm	4	$ & $	-0.10	$\\
1543254	&	2456103.9324	&	1054	& $	7.0	\pm	4	$ & $	1.75	$& $	0.9	\pm	4	$ & $	0.22	$\\
1543258	&	2456103.9369	&	1001	& $	0.8	\pm	4	$ & $	0.20	$& $	0.1	\pm	4	$ & $	0.03	$\\
1543262	&	2456103.9414	&	983	& $	3.8	\pm	4	$ & $	0.91	$& $	3.1	\pm	4	$ & $	0.75	$\\
1543266	&	2456103.9459	&	1066	& $	-3.3	\pm	4	$ & $	-0.88	$& $	1.0	\pm	4	$ & $	0.27	$\\
1543270	&	2456103.9505	&	1015	& $	0.4	\pm	4	$ & $	0.11	$& $	-1.3	\pm	4	$ & $	-0.32	$\\
1543274	&	2456103.9554	&	1047	& $	-3.2	\pm	4	$ & $	-0.84	$& $	3.2	\pm	4	$ & $	0.84	$\\
1543278	&	2456103.9599	&	1059	& $	-1.3	\pm	4	$ & $	-0.34	$& $	-3.6	\pm	4	$ & $	-0.94	$\\
1543282	&	2456103.9645	&	1086	& $	-0.7	\pm	4	$ & $	-0.20	$& $	-2.2	\pm	4	$ & $	-0.59	$\\
1543286	&	2456103.9690	&	1125	& $	-2.4	\pm	4	$ & $	-0.66	$& $	4.3	\pm	4	$ & $	1.18	$\\
1543290	&	2456103.9735	&	1147	& $	-5.7	\pm	4	$ & $	-1.62	$& $	-0.2	\pm	4	$ & $	-0.05	$\\
1543294	&	2456103.9781	&	1093	& $	0.9	\pm	4	$ & $	0.25	$& $	-2.9	\pm	4	$ & $	-0.79	$\\
1543298	&	2456103.9826	&	1142	& $	2.7	\pm	4	$ & $	0.74	$& $	-3.5	\pm	4	$ & $	-0.95	$\\
1543302	&	2456103.9871	&	1124	& $	-0.7	\pm	4	$ & $	-0.19	$& $	1.5	\pm	4	$ & $	0.41	$\\
1543306	&	2456103.9916	&	1060	& $	2.3	\pm	4	$ & $	0.61	$& $	-4.5	\pm	4	$ & $	-1.19	$\\
1543486	&	2456104.8906	&	960	& $	1.4	\pm	4	$ & $	0.33	$& $	-0.3	\pm	4	$ & $	-0.06	$\\
1543490	&	2456104.8951	&	1064	& $	-2.6	\pm	4	$ & $	-0.67	$& $	-1.4	\pm	4	$ & $	-0.35	$\\
1543494	&	2456104.8996	&	972	& $	-1.6	\pm	4	$ & $	-0.38	$& $	-5.0	\pm	4	$ & $	-1.16	$\\
1543498	&	2456104.9041	&	928	& $	2.3	\pm	4	$ & $	0.53	$& $	-1.0	\pm	4	$ & $	-0.23	$\\
1543502	&	2456104.9087	&	953	& $	-5.8	\pm	4	$ & $	-1.41	$& $	0.7	\pm	4	$ & $	0.17	$\\
1543506	&	2456104.9132	&	1004	& $	4.9	\pm	4	$ & $	1.21	$& $	-5.6	\pm	4	$ & $	-1.37	$\\
1543510	&	2456104.9178	&	944	& $	4.3	\pm	4	$ & $	1.00	$& $	-1.2	\pm	4	$ & $	-0.27	$\\
1543514	&	2456104.9223	&	1035	& $	-0.8	\pm	4	$ & $	-0.21	$& $	0.7	\pm	4	$ & $	0.17	$\\
1543518	&	2456104.9268	&	945	& $	1.0	\pm	4	$ & $	0.25	$& $	7.7	\pm	5	$ & $	1.69	$\\
1543522	&	2456104.9318	&	927	& $	-1.4	\pm	4	$ & $	-0.33	$& $	0.1	\pm	4	$ & $	0.01	$\\
1543526	&	2456104.9363	&	914	& $	-0.1	\pm	4	$ & $	-0.03	$& $	0.5	\pm	4	$ & $	0.11	$\\
1543530	&	2456104.9409	&	939	& $	2.2	\pm	4	$ & $	0.53	$& $	9.4	\pm	4	$ & $	2.21	$\\
1543534	&	2456104.9454	&	882	& $	-7.4	\pm	5	$ & $	-1.62	$& $	-1.1	\pm	5	$ & $	-0.24	$\\
1543538	&	2456104.9499	&	974	& $	0.4	\pm	4	$ & $	0.11	$& $	3.0	\pm	4	$ & $	0.71	$\\
1543542	&	2456104.9545	&	978	& $	-4.2	\pm	4	$ & $	-1.02	$& $	-5.9	\pm	4	$ & $	-1.43	$\\
1543546	&	2456104.9590	&	959	& $	2.7	\pm	4	$ & $	0.63	$& $	-1.4	\pm	4	$ & $	-0.32	$\\
1543550	&	2456104.9635	&	998	& $	-2.6	\pm	4	$ & $	-0.62	$& $	-2.1	\pm	4	$ & $	-0.49	$\\
1543554	&	2456104.9681	&	1008	& $	0.4	\pm	4	$ & $	0.10	$& $	-4.9	\pm	4	$ & $	-1.17	$\\
1543558	&	2456104.9731	&	1057	& $	-5.3	\pm	4	$ & $	-1.33	$& $	-1.1	\pm	4	$ & $	-0.27	$\\
1543562	&	2456104.9776	&	1021	& $	-1.1	\pm	4	$ & $	-0.28	$& $	-0.2	\pm	4	$ & $	-0.06	$\\
1543566	&	2456104.9821	&	1017	& $	-2.7	\pm	4	$ & $	-0.69	$& $	2.7	\pm	4	$ & $	0.67	$\\
1543570	&	2456104.9866	&	1014	& $	6.6	\pm	4	$ & $	1.69	$& $	3.4	\pm	4	$ & $	0.86	$\\
1543574	&	2456104.9911	&	975	& $	-1.6	\pm	4	$ & $	-0.41	$& $	4.1	\pm	4	$ & $	1.02	$\\
1543578	&	2456104.9957	&	1025	& $	-4.3	\pm	4	$ & $	-1.11	$& $	3.1	\pm	4	$ & $	0.77	$\\
1543582	&	2456105.0002	&	1027	& $	-8.4	\pm	4	$ & $	-2.13	$& $	6.9	\pm	4	$ & $	1.73	$\\
1543586	&	2456105.0047	&	1022	& $	2.1	\pm	4	$ & $	0.53	$& $	-6.3	\pm	4	$ & $	-1.56	$\\
1543590	&	2456105.0092	&	931	& $	-1.7	\pm	4	$ & $	-0.39	$& $	-0.7	\pm	5	$ & $	-0.16	$\\
1543594	&	2456105.0144	&	943	& $	4.4	\pm	4	$ & $	1.05	$& $	-10.0	\pm	4	$ & $	-2.32	$\\
1543598	&	2456105.0190	&	1020	& $	1.7	\pm	4	$ & $	0.42	$& $	1.0	\pm	4	$ & $	0.26	$\\
1543602	&	2456105.0235	&	886	& $	-3.9	\pm	5	$ & $	-0.83	$& $	0.9	\pm	5	$ & $	0.19	$\\
		  \hline
\end{tabular}
\end{center}
\end{table}

\begin{table}
\addtolength{\tabcolsep}{-1pt}
\contcaption{Spectropolarimetric observations.}
\begin{center}
\begin{tabular}{ccrrrrrrrrrrrr}
\hline
\#	&	HJD	&	S/N	& 	$\bz$			&	$z_V$	&	$\nz$			&	$z_N$	\\
\hline		
1543606	&	2456105.0280	&	837	& $	1.6	\pm	5	$ & $	0.34	$& $	0.3	\pm	5	$ & $	0.06	$\\
1543610	&	2456105.0325	&	824	& $	6.0	\pm	5	$ & $	1.26	$& $	1.3	\pm	5	$ & $	0.27	$\\
1543614	&	2456105.0371	&	837	& $	-1.7	\pm	5	$ & $	-0.37	$& $	-7.2	\pm	5	$ & $	-1.54	$\\
1543618	&	2456105.0416	&	770	& $	-2.5	\pm	5	$ & $	-0.49	$& $	2.9	\pm	5	$ & $	0.56	$\\
1543622	&	2456105.0461	&	840	& $	3.0	\pm	5	$ & $	0.63	$& $	-0.8	\pm	5	$ & $	-0.17	$\\
1543626	&	2456105.0506	&	858	& $	-0.5	\pm	5	$ & $	-0.10	$& $	-5.9	\pm	5	$ & $	-1.23	$\\
1543988	&	2456105.9391	&	1209	& $	1.3	\pm	3	$ & $	0.39	$& $	2.0	\pm	3	$ & $	0.59	$\\
1543992	&	2456105.9437	&	1259	& $	-1.9	\pm	3	$ & $	-0.57	$& $	2.9	\pm	3	$ & $	0.87	$\\
1543996	&	2456105.9482	&	1268	& $	4.3	\pm	3	$ & $	1.30	$& $	3.0	\pm	3	$ & $	0.93	$\\
1544000	&	2456105.9527	&	1168	& $	-0.7	\pm	3	$ & $	-0.20	$& $	-5.1	\pm	3	$ & $	-1.49	$\\
1544004	&	2456105.9572	&	1066	& $	0.2	\pm	4	$ & $	0.07	$& $	0.5	\pm	4	$ & $	0.14	$\\
1544012	&	2456105.9663	&	1138	& $	-1.0	\pm	4	$ & $	-0.28	$& $	3.4	\pm	4	$ & $	0.96	$\\
1544016	&	2456105.9708	&	1068	& $	-2.2	\pm	4	$ & $	-0.58	$& $	1.1	\pm	4	$ & $	0.29	$\\
1544020	&	2456105.9753	&	1130	& $	0.6	\pm	4	$ & $	0.18	$& $	5.7	\pm	4	$ & $	1.59	$\\
1544024	&	2456105.9802	&	992	& $	-2.3	\pm	4	$ & $	-0.58	$& $	-6.2	\pm	4	$ & $	-1.53	$\\
1544028	&	2456105.9847	&	1057	& $	-0.6	\pm	4	$ & $	-0.15	$& $	-1.4	\pm	4	$ & $	-0.35	$\\
1544032	&	2456105.9893	&	1152	& $	-0.2	\pm	4	$ & $	-0.06	$& $	-3.7	\pm	4	$ & $	-1.00	$\\
1544036	&	2456105.9938	&	1094	& $	-6.1	\pm	4	$ & $	-1.67	$& $	-5.4	\pm	4	$ & $	-1.48	$\\
1544040	&	2456105.9983	&	1124	& $	1.0	\pm	4	$ & $	0.29	$& $	0.2	\pm	4	$ & $	0.05	$\\
1544044	&	2456106.0029	&	1164	& $	1.9	\pm	3	$ & $	0.56	$& $	3.5	\pm	3	$ & $	1.04	$\\
1544048	&	2456106.0074	&	1117	& $	2.4	\pm	4	$ & $	0.65	$& $	0.1	\pm	4	$ & $	0.02	$\\
1544052	&	2456106.0119	&	1084	& $	0.6	\pm	4	$ & $	0.17	$& $	6.7	\pm	4	$ & $	1.75	$\\
1544056	&	2456106.0165	&	1014	& $	-10.8	\pm	4	$ & $	-2.67	$& $	-1.6	\pm	4	$ & $	-0.39	$\\
1544060	&	2456106.0214	&	1147	& $	2.1	\pm	4	$ & $	0.59	$& $	1.2	\pm	4	$ & $	0.33	$\\
1544064	&	2456106.0260	&	1174	& $	0.2	\pm	4	$ & $	0.06	$& $	-2.3	\pm	4	$ & $	-0.66	$\\
1544068	&	2456106.0305	&	1140	& $	-2.3	\pm	4	$ & $	-0.63	$& $	0.8	\pm	4	$ & $	0.21	$\\
1544072	&	2456106.0350	&	1154	& $	3.1	\pm	4	$ & $	0.87	$& $	1.9	\pm	4	$ & $	0.53	$\\
1544076	&	2456106.0395	&	1086	& $	-4.6	\pm	4	$ & $	-1.29	$& $	-4.1	\pm	4	$ & $	-1.09	$\\
1544080	&	2456106.0441	&	1134	& $	4.1	\pm	4	$ & $	1.14	$& $	-1.2	\pm	4	$ & $	-0.32	$\\
1544084	&	2456106.0486	&	1134	& $	1.7	\pm	4	$ & $	0.47	$& $	4.4	\pm	4	$ & $	1.19	$\\
1544088	&	2456106.0531	&	1039	& $	-2.1	\pm	4	$ & $	-0.55	$& $	-5.0	\pm	4	$ & $	-1.27	$\\
1544092	&	2456106.0576	&	1112	& $	-4.3	\pm	4	$ & $	-1.18	$& $	0.8	\pm	4	$ & $	0.21	$\\
1544447	&	2456106.9078	&	578	& $	-6.8	\pm	7	$ & $	-0.98	$& $	0.2	\pm	7	$ & $	0.03	$\\
1544451	&	2456106.9124	&	569	& $	9.8	\pm	7	$ & $	1.40	$& $	-7.0	\pm	7	$ & $	-1.01	$\\
1544455	&	2456106.9169	&	568	& $	9.0	\pm	7	$ & $	1.27	$& $	6.1	\pm	7	$ & $	0.84	$\\
1544459	&	2456106.9214	&	507	& $	-13.5	\pm	8	$ & $	-1.74	$& $	-10.2	\pm	8	$ & $	-1.31	$\\
1544463	&	2456106.9259	&	508	& $	-1.3	\pm	8	$ & $	-0.16	$& $	2.0	\pm	8	$ & $	0.25	$\\
1544467	&	2456106.9305	&	459	& $	-6.9	\pm	8	$ & $	-0.81	$& $	-14.4	\pm	8	$ & $	-1.70	$\\
1544471	&	2456106.9350	&	510	& $	-11.7	\pm	8	$ & $	-1.50	$& $	3.2	\pm	8	$ & $	0.40	$\\
1544475	&	2456106.9395	&	365	& $	-11.6	\pm	11	$ & $	-1.05	$& $	14.3	\pm	11	$ & $	1.26	$\\
1544479	&	2456106.9441	&	501	& $	3.9	\pm	8	$ & $	0.48	$& $	-5.8	\pm	8	$ & $	-0.69	$\\
1544483	&	2456106.9490	&	390	& $	-7.7	\pm	10	$ & $	-0.77	$& $	-8.1	\pm	10	$ & $	-0.79	$\\
1544487	&	2456106.9536	&	327	& $	-6.2	\pm	12	$ & $	-0.53	$& $	4.8	\pm	12	$ & $	0.40	$\\
1544491	&	2456106.9581	&	384	& $	-3.5	\pm	10	$ & $	-0.34	$& $	3.6	\pm	10	$ & $	0.35	$\\
1544495	&	2456106.9626	&	319	& $	-0.9	\pm	12	$ & $	-0.07	$& $	9.4	\pm	13	$ & $	0.75	$\\
1544499	&	2456106.9672	&	319	& $	6.3	\pm	12	$ & $	0.51	$& $	9.7	\pm	13	$ & $	0.77	$\\
1544503	&	2456106.9717	&	611	& $	-4.2	\pm	7	$ & $	-0.64	$& $	-6.5	\pm	6	$ & $	-1.00	$\\
1544507	&	2456106.9762	&	702	& $	1.9	\pm	5	$ & $	0.35	$& $	-2.8	\pm	6	$ & $	-0.50	$\\
1544511	&	2456106.9807	&	626	& $	5.4	\pm	6	$ & $	0.86	$& $	-1.3	\pm	6	$ & $	-0.20	$\\
1544515	&	2456106.9853	&	667	& $	-0.8	\pm	6	$ & $	-0.15	$& $	-1.4	\pm	6	$ & $	-0.24	$\\
1544519	&	2456106.9904	&	732	& $	1.3	\pm	5	$ & $	0.24	$& $	3.7	\pm	5	$ & $	0.68	$\\
1544523	&	2456106.9949	&	711	& $	5.8	\pm	5	$ & $	1.08	$& $	-10.8	\pm	5	$ & $	-1.98	$\\
1544527	&	2456106.9994	&	453	& $	8.8	\pm	9	$ & $	1.01	$& $	1.5	\pm	9	$ & $	0.16	$\\
1544730	&	2456107.9126	&	1188	& $	2.9	\pm	3	$ & $	0.86	$& $	0.2	\pm	4	$ & $	0.05	$\\
1544734	&	2456107.9172	&	1083	& $	-1.5	\pm	4	$ & $	-0.42	$& $	5.4	\pm	4	$ & $	1.45	$\\
1544738	&	2456107.9217	&	1082	& $	0.1	\pm	4	$ & $	0.04	$& $	0.1	\pm	4	$ & $	0.03	$\\
1544742	&	2456107.9262	&	1104	& $	-4.5	\pm	4	$ & $	-1.23	$& $	2.4	\pm	4	$ & $	0.64	$\\
		  \hline
\end{tabular}
\end{center}
\end{table}

\begin{table}
\addtolength{\tabcolsep}{-1pt}
\contcaption{Spectropolarimetric observations.}
\begin{center}
\begin{tabular}{ccrrrrrrrrrrrr}
\hline
\#	&	HJD	&	S/N	& 	$\bz$			&	$z_V$	&	$\nz$			&	$z_N$	\\
\hline	
1544746	&	2456107.9307	&	969	& $	2.5	\pm	4	$ & $	0.60	$& $	0.2	\pm	4	$ & $	0.04	$\\
1544750	&	2456107.9353	&	1099	& $	-4.8	\pm	4	$ & $	-1.28	$& $	1.4	\pm	4	$ & $	0.36	$\\
1544754	&	2456107.9398	&	1110	& $	3.6	\pm	4	$ & $	0.97	$& $	1.7	\pm	4	$ & $	0.44	$\\
1544758	&	2456107.9443	&	1145	& $	0.6	\pm	4	$ & $	0.18	$& $	3.6	\pm	4	$ & $	1.01	$\\
1544762	&	2456107.9489	&	1175	& $	-1.8	\pm	3	$ & $	-0.51	$& $	-0.3	\pm	4	$ & $	-0.08	$\\
1544766	&	2456107.9541	&	989	& $	-6.1	\pm	4	$ & $	-1.50	$& $	3.7	\pm	4	$ & $	0.90	$\\
1544770	&	2456107.9586	&	1033	& $	2.7	\pm	4	$ & $	0.68	$& $	3.3	\pm	4	$ & $	0.84	$\\
1544774	&	2456107.9631	&	1059	& $	8.5	\pm	4	$ & $	2.24	$& $	-1.2	\pm	4	$ & $	-0.31	$\\
1544778	&	2456107.9676	&	1073	& $	1.0	\pm	4	$ & $	0.28	$& $	1.0	\pm	4	$ & $	0.26	$\\
1544782	&	2456107.9722	&	1073	& $	-5.1	\pm	4	$ & $	-1.36	$& $	1.3	\pm	4	$ & $	0.36	$\\
1544786	&	2456107.9767	&	1135	& $	1.8	\pm	4	$ & $	0.51	$& $	7.5	\pm	4	$ & $	2.05	$\\
1544790	&	2456107.9812	&	1190	& $	0.5	\pm	3	$ & $	0.14	$& $	-1.9	\pm	3	$ & $	-0.55	$\\
1544794	&	2456107.9858	&	1182	& $	-0.2	\pm	4	$ & $	-0.06	$& $	1.7	\pm	4	$ & $	0.47	$\\
1544798	&	2456107.9903	&	1136	& $	-3.8	\pm	4	$ & $	-1.05	$& $	0.3	\pm	4	$ & $	0.07	$\\
\hline
\end{tabular}
\end{center}
\end{table}

\end{document}